\documentclass[superscriptaddress,twocolumn,showpacs,
amssymb,amsmath,nobibnotes,aps,prd,
nofootinbib]{revtex4-1}
\pdfoutput=1
\usepackage{graphicx,subfigure,bm,color,psfrag,hyperref}
\usepackage{amsfonts}
\usepackage{lipsum}
\usepackage{multirow}
\usepackage{booktabs}
\usepackage{subcaption}
\usepackage{mathtools}
\usepackage{verbatim}
\usepackage[normalem]{ulem}
\usepackage[dvipsnames]{xcolor}
\hypersetup{colorlinks,linkcolor={blue},citecolor={red},urlcolor={magenta}}

\usepackage{color}

\begin{document}

\title{Non-Commutative Fluid as an Alternative Driver of Cosmic Acceleration: Confronting DESI Observations}

\author{Raj Kumar Das}
\email{raj1996cool@gmail.com}
\email{rajdas\_r@isical.ac.in}
\affiliation{Physics and Applied Mathematics Unit, Indian Statistical Institute, 203 B. T. Road, Kolkata 700108, India}

\author{Arpan Krishna Mitra}
\email{arpankmitra@gmail.com}
\email{arpankmitra@cuap.edu.in}
\affiliation{Space Science and Technology
Central University of Andhra Pradesh,PMW3+M26, Neelampalli Rd, Janthalur, Andhra Pradesh 515701}

\begin{abstract}

We present a novel formulation for the Hubble parameter derived from Newtonian cosmology, incorporating non-commutative fluid dynamics through a deformed Poisson bracket structure. This approach introduces a new cosmological parameter, denoted by $\sigma$, which emerges naturally from the underlying non-commutative framework. It gives rise to a source term in the background fluid continuity equation, thereby leading to an apparent type of matter creation picture through the resulting non-conservation. Remarkably, the resulting Hubble function accounts for the observed accelerated expansion of the universe without invoking any external dark energy component or cosmological constant. Instead, the parameter $\sigma$ effectively serves as the driver of acceleration. We further examine the observational constraints on $\sigma$ using current cosmological data, including the recent Dark Energy Spectroscopic Instrument(DESI) dataset, demonstrating its viability as an alternative explanation for late-time cosmic acceleration within a non-commutative cosmological model.
\end{abstract}

\maketitle
\section{Introduction}
 The experiments involving type Ia supernovae (SNe Ia) data strongly suggest a late-time acceleration of the universe ~\cite{SupernovaSearchTeam:1998fmf,SupernovaCosmologyProject:1998vns}. Current observations dictate the existence of a Friedmann-Lemaitre cosmology with zero curvature, where CDM and baryons are the dominant components in the matter sector at present and a cosmological constant $\Lambda$. In the early phase, both matter and radiation were dominant, leading to decelerated expansion of the universe. {The expansion of the universe causes the energy densities of the matter and radiation sectors to dilute. Consequently, it becomes necessary to introduce vacuum energy, represented by the Cosmological Constant ($\Lambda$) ~\cite{Js:2013pub}, to account for the accelerating expansion. This component has become dominant in the recent cosmic epoch, driving the observed late-time acceleration.} For this reason, the $\Lambda$CDM model is considered a well-established vanilla model to explain cosmic acceleration and large-scale structure of the universe. However, the model has some serious issues, including tension anomalies~\cite{Pan:2023mie,Anchordoqui_2021,DiValentino:2021izs,DiValentino:2021zxy,Anchordoqui:2021gji,Li:2019yem,Yang:2021eud}, the coincidence problem~\cite{Zlatev:1998tr}, and the cosmological constant problem~\cite{Weinberg:1988cp}, among others. These issues are the key motivation for the development of alternative cosmological models that address these drawbacks. A plethora of models have been proposed, approaching the problem from different perspectives, including phenomenological models and extended theoretical frameworks such as Modified Gravity, Scalar Field Models or any other non-standard theories. 

In general, the ~$\Lambda$CDM model consists of two components, a "Cosmological Constant" ~($\Lambda$) playing the role of "Dark Energy" and the "Cold Dark Matter" (CDM), where the term "cold" refers to the non-relativistic nature of dark matter.  As mentioned previously, several alternative proposals try to explain the accelerating expansion phase. "Matter Creation Cosmology" is one of the recent theories among them ~\cite{Steigman:2008bc,Lima:2009ic,Lima:2015xpa,Basilakos:2010yp,Alcaniz:1999hu,Freaza:2002ic,Halder:2025eze}.

The concept of matter creation deals with the creation of cold dark matter(CCDM) due to gravitationally induced particle(non-relativistic species) production. This particle production may lead to the required negative pressure, so it can be an alternative to "Dark Energy". This behaviour of driving the cosmic acceleration through particle production phenomena has also been presented in ~\cite{Zimdahl:2000zm}.
 In the matter creation scenario, there is no need for a cosmological constant or any quintessence-type component.
There are several perspectives on how to outline the concept of matter creation. In this work, we have approached matter creation through a noncommutativity-corrected fluid term in Newtonian cosmology.

'Noncommutative (NC) spacetime effects in ideal fluid dynamics' is an area that has experienced some
recent interest~\cite{Das:2016hmc,Das:2017ytv,Das:2018puw,Ren:2018rku}. In an earlier work~ \cite{Mitra:2018ezo}, we have dealt with our
extended NC fluid variable algebra and a subsequent study of the generalised continuity  equation. This continuity equation reveals the non-conservative behaviour of background matter density, stemming from a source term that emerges as a direct manifestation of NC effects. These effects modify the standard continuity equation, signalling an effective creation or exchange of matter-energy within the cosmic background.  We have provided an outline of the effects on
cosmological principles induced by the NC-modified fluid  dynamics. 

By implementing the extended NC fluid dynamic equations, we have derived a modified Friedmann equation (FE) in which the extended NC fluid affects the matter sector and generates an effective curvature component. When the extended NC terms vanish, the FE simplifies to a non-interacting matter-curvature scenario. No additional effective cosmological constant arises. Instead, it modifies the matter and curvature components, influencing their individual equations of state. In this way, we have established a connection between extended NC fluids and the matter creation scenario.

This paper is organised as follows: in section-\ref{FE}, we have discussed the theoretical foundation of our model, the background cosmic flow dynamics. The effects of NC extension are explained in section-\ref{BC}. The details of the dataset used for our analysis are provided in section-\ref{data}. The observational constraints regarding our model are discussed in section-\ref{RE}. Finally, in section-\ref{sum}, we summarise and discuss about the future extension of our work.

\section{Friedmann equation motivated from Non-Commutative Cosmology }
\label{FE}
{The very first step to deal with NC fluid is the scheme one follows to introduce the NC effect in the fluid dynamic equations. We have introduced NC algebra in Lagrangian (discrete) fluid degrees of freedom, which subsequently percolates to the Euler (field) fluid degrees of freedom\cite{Mitra:2018ezo}. { The modified NC algebra, which is the key term of this work has been defined in term of poission brackets;
\begin{align}
\label{2a}
\{X_{i}({\bf{x}}), X_{j}({\bf{y}})\} = \frac{\theta_{ij}}{\rho_{0}}\delta ({\bf{x}}-{\bf{y}}),~ \nonumber \\
\{\dot{X_{i}}({\bf{x}}), X_{j}({\bf{y}})\}=\frac{1}{\rho_{0}}(\delta_{ij} +\sigma_{ij})\delta ({\bf{x}}-{\bf{y}}),~ \nonumber \\\{\dot{X_{i}({\bf{x}})}, \dot{X_{j}}({\bf{y}})\}=0.
\end{align}
$\theta_{ij}$ is the constant antisymmetric ($\theta_{ij} = -\theta_{ji}$) Non-Commutative (NC) parameter tensor. The parameter $\sigma_{ij}$ is an NC parameter introduced in the Poisson bracket of the position and momentum variables, serving more fundamentally as a deformation parameter for subsequently breaking the inherent symmetry \cite{Mitra:2018ezo}.
} {Unlike the diagonal Kronecker delta $\delta_{ij}$, the inclusion of $\sigma_{ij}$ introduces off-diagonal terms such as $\sigma_{12}$, $\sigma_{13}$, and $\sigma_{23}$. Physically, this indicates a cross-coupling where momentum changes in one direction directly influence the position in orthogonal directions.}

{ This generalization of adding $\sigma_{ij}$ has been extensively studied as $\alpha$-star deformation Poisson algebra through the implementation of the ``Moyal Product'' \cite{MARCIAL20103608, Abreu_2025, Djemai:2003kd}. In these literatures, a perspective known as ``Quantization by deformation'' is typically employed \cite{Djemai:2003yn}. For the first time in \cite{Mitra:2018ezo}, the deformation effect was studied from a perturbative perspective within the framework of theoretical Newtonian cosmology. 

In the aforementioned literature, there exists a pure mathematical and physical connection between $\sigma_{ij}$ and $\theta_{ij}$(The non-commutative parameters $\theta_{ij}$, $\beta_{ij}$, and $\sigma_{ij}$ are not independent; they are constrained by the Jacobi identities to ensure the structural consistency of the deformed phase-space algebra. Specifically, the identities $\{x_i, \{x_j, p_k\}\} + \text{cyclic} = 0$ and $\{p_i, \{p_j, x_k\}\} + \text{cyclic} = 0$ lead to a set of consistency conditions. These relations imply that the spatial "fuzziness" ($\theta$), the momentum-space deformation ($\beta$), and the cross-coupling term ($\sigma$) must evolve or be scaled in a mutually dependent manner to maintain a closed algebraic structure. In the perturbative limit where $|\sigma| < 1$, these constraints ensure that the symplectic manifold remains well-defined.  Also this effects are clearly mebtioned in \cite{Mitra:2018ezo} in terms of velocity density modified algebra); however, in the present study, there is no such physical connectivity between them. Here, $\theta_{ij}$ primarily denotes the fuzziness of the space, rescaled by the background density $\rho_{0}$, as is common in the form $\delta_{ij} + \sigma_{ij}$.} \\

Based on this approach, the modified Euler dynamics with NC-terms will be explained through the following equations of the density and velocity parts,

\begin{align}
\label{2an}
\dot{\rho}=\{\rho , H\}=-\partial_{i}(\rho v_{i})-\sigma_{ij}\partial_{j}(\rho v_{i}) \nonumber \\ =-\partial_{i}(\rho v_{i} +\sigma_{ji}\rho v_{j}),
\end{align}
\begin{align}   
\label{2al}
\dot{v_{k}}=\{v_{k}, H\}=-v_{i}\partial_{i}v_{k} - \sigma_{ij}v_{i}\partial_{j}v_{k}-  \nonumber \\  \partial_{k}V'(\rho)- \sigma_{kj}\partial_{j}V'(\rho)+\theta_{ji}\partial_{i}V'(\rho)
\partial_{j}v_{k}.
\end{align}
{ The equations above describe the modified Euler dynamics within the non-commutative framework. These expressions incorporate corrections arising from the non-commutative parameters $\sigma_{ij}$ and $\theta_{ij}$, which account for the fundamental modifications to the fluid's evolution.}
Here $H$ is the Hamiltonian, defined as $H=\int dV(\frac{1}{2}\rho v^{2} +V(\rho))$ and   we always consider the comoving frame in the current FLRW cosmological scenario. Starting from the relation ${\bf r}(t) = a(t){\bf x}$, where $a(t)$ is the cosmic scale factor and ${\bf x}$ is the comoving coordinate, we obtain the physical velocity as $\dot{\bf r} = {\bf u} = \dot{a} {\bf x} + a \dot{\bf x}$. Here, the term ${\bf v} = a \dot{\bf x}$ represents the peculiar velocity, which arises in perturbative scenarios.

It is important to consider the transformation of space and time derivatives between the laboratory (physical) and comoving frames. These transformations take the form:
$$
\frac{\partial}{\partial {\bf r}} = \frac{1}{a} \frac{\partial}{\partial {\bf x}}, \qquad
\left.\frac{\partial}{\partial t}\right|_{\bf r} = \left.\frac{\partial}{\partial t}\right|_{\bf x} - \frac{\dot{a}}{a} ({\bf x} \cdot \nabla_{\bf x}).
$$
A more detailed discussion of these topics can be found in \cite{Mitra:2018ezo}. In this study, all calculations are performed within the Newtonian framework, as the analysis proceeds based on a perturbative fluid dynamics approach. The Newtonian framework is specifically chosen because incorporating Non-Commutative (NC) corrections into an effective energy-momentum tensor within a full general relativistic perspective would introduce significant  conceptual complexities. Consequently, we focus on the Newtonian limit to maintain a tractable and clear perturbative treatment of the NC effects. Newtonian cosmology has some limitations. It cannot fully describe the radiation era, CMB physics, or super-horizon scales. However, it is a simple and useful way to study late-time cosmic evolution. In this paper, we use this framework to show how non-commutative effects change the expansion of the universe. This approach is common, and several other studies also use Newtonian cosmological frameworks ~ \cite{Das:2018puw,Ma:2018sts,Buchert:1995fz,Buchert:1999pq}. 
Using the framework of noncommutative (NC)-based Euler dynamics, along with all the criteria mentioned above, we obtain the following modified continuity equations:

\begin{equation}
\label{aj3}
\dot{\rho}+ 3\frac{\dot{a}}{a}\rho +\partial_{i}(\rho v_{i})+\frac{\sigma_{ij}}{a}\partial_{j}(\rho \dot{a}x_{i}+\rho v_{i})=0.
\end{equation}

{Expanding Equation~\eqref{aj3}, we derive the continuity relation for the background component. We decompose the density as $\rho(\mathbf{x},t) = \rho_0(t) + \delta\rho(\mathbf{x},t)$. By vanishing the peculiar velocity components to isolate the background evolution, we obtain:
\begin{equation}
    \dot{\rho}_{0} + 3\frac{\dot{a}}{a}\rho_{0} + \frac{1}{a}\sigma_{ij}\partial_{j}(\rho_{0}\dot{a}x_{i}) = 0.
\end{equation}

Since $\rho_0$ and $\dot{a}$ are independent of spatial coordinates, the third term simplifies as follows:
\begin{equation}
    \sigma_{ij}\partial_{j}(\rho_{0}\dot{a}x_{i}) = \rho_0\dot{a} \sigma_{ij}\partial_j(x_i) = \rho_0\dot{a} \sigma_{ij}\delta_{ij}.
\end{equation}

By applying the trace condition $\delta_{ij}\sigma_{ij} = \sigma$ and substituting the Hubble parameter $H = \frac{\dot{a}}{a}$, we arrive at the modified background evolution equation:
\begin{equation}
    \dot{\rho}_0 + 3H \rho_0 = -\sigma H \rho_0.
    \label{bc}
\end{equation}}

{In the presence of a Non-Commutative (NC) generalized fluid, the background matter density $\rho_0$ is not conserved, as demonstrated in equation~\eqref{bc}. Here, $\sigma \equiv \delta_{ij}\sigma_{ij}$ represents the trace of the NC parameter tensor $\sigma_{ij}$. The term $-\sigma H \rho_{0}$ functions as an effective source term, mimicking the creation of matter particles or energy transfer processes observed in specific cosmological scenarios. 

However, it is important to clarify that this does not necessarily imply physical particle production or decay in the traditional sense, as no external field or source is introduced. Instead, this represents an ``apparent'' matter creation resulting from the modification of the Poisson algebra. The deformation of the symplectic structure, caused by $\sigma_{ij}$, alters the phase-space geometry and the associated volume measure. Consequently, the background expansion interacts with this deformed geometry to produce a shift in the density evolution, which can be interpreted as a geometric energy exchange between the spacetime manifold and the fluid.}

The background matter fluid $\rho_0$ follows a modified continuity equation that incorporates the NC-corrected equation of state, given by:
\begin{equation}
    \dot{\rho_0}+H(3+\sigma)\rho_0 = 0 \implies \rho_0 = \bar\rho_0 a^{-(3+\sigma)}
    \label{eqn-continuty}
\end{equation}

The dynamical equations for an ideal fluid are modified by these non-commutative (NC) effects. By considering the NC-corrected Euler equation containing the velocity term in Eq.~\eqref{2al}, one can derive the explicit form of the Friedmann acceleration equation. { The detailed derivation of this equation, including the NC-correction term, is provided in Ref.~\cite{Mitra:2018ezo} and is expressed as}:
\begin{equation}
\frac{\ddot{a}}{a}+\frac{4\pi G\rho_0}{3}+\sigma H^2 = 0
\label{eqn-acceleration}
\end{equation}
In this context, $\rho_0$ denotes the background density of the matter sector within the cosmic fluid, while the term $\sigma H^2$ represents the NC correction. By substituting the acceleration term $\frac{\ddot{a}}{a} = \dot H + H^2$ into Equation~\eqref{eqn-acceleration}, we obtain:
\begin{equation}
    \frac{dH}{dt}+H^2(1+\sigma)+\frac{4\pi G\rho_0}{3} = 0.
    \label{eqn-need2}
\end{equation}
Utilizing the relation $\dot{a}=aH$, Equation~\eqref{eqn-need2} becomes:
\begin{equation}
    \frac{dH^2}{da}+ H^2\frac{2(1+\sigma)}{a} = - \frac{8 \pi G }{3a}\rho_0.
    \label{eqn-new}
\end{equation}

Applying the method of the integrating factor, we arrive at the following expression:
\begin{equation}
    \frac{d}{da}\bigg[H^2 a^{2(1+\sigma)}\bigg] = - \frac{8 \pi G }{3a}a^{2(1+\sigma)}\rho_0.
\end{equation}

{ {By combining the NC-extended continuity Equation~\eqref{eqn-continuty} with Equation~\eqref{eqn-new}, we proceed with the integration:
\begin{equation}
    H^2 a^{2(1+\sigma)} = - \frac{8 \pi G \bar\rho_0}{3}\int a^{-(3+\sigma)} a^{(2+2\sigma)-1} da
\end{equation}
Evaluating the integral yields:
\begin{equation}
     H^2 a^{2(1+\sigma)} = - \frac{8 \pi G \bar\rho_0}{3} \bigg( \frac{a^{\sigma-1}}{\sigma-1}\bigg)+ C =  \frac{8 \pi G \bar\rho_0}{3} \bigg( \frac{a^{(\sigma-1)}}{(1-\sigma)}\bigg)+C
\end{equation}
where $C$ is the integration constant. Following algebraic manipulation and applying the boundary condition $H(a=a_0=1) = H_0$, we find:
\begin{equation}
    C = H_0^2 -  \frac{8 \pi G \bar\rho_0}{3} \bigg( \frac{1}{1-\sigma}\bigg)
\end{equation}

Finally, by substituting the value of $C$ and employing the redshift relation $z = \frac{1}{a} - 1$, we derive the expression for the Hubble function:
\begin{equation}
    H^2 = H^2_{0}\Bigg[\frac{\Omega_{0}}{1-\sigma}(1+z)^{(3+\sigma)} + (1- \frac{\Omega_{0}}{1-\sigma})(1+z)^{(2\sigma+2)}\Bigg].
    \label{Hubble}
\end{equation}}
Here, $H_0$ is the present-day Hubble parameter and the density parameter $\Omega_0$ is defined as:
\begin{equation*}
\Omega_0 = \frac{\bar\rho_0}{\rho_{crit}} = \frac{\bar\rho_0}{\frac{3 H_0^2}{8 \pi G}},
\end{equation*}}
which is derived from the present-day background fluid density $\bar{\rho_0}$.
 From \eqref{Hubble}, {it is apparent that the density parameter and the equation of state parameter receive significant modification in the presence of  $\sigma$. As $\sigma$ appears in the fluid dynamical equations as NC correction term, it should follow the theoretical bound $-1 <\sigma < 1$. { It is important to note that within the modified Poisson bracket, the non-commutative parameter $\sigma_{ij}$ is added to the Kronecker delta $\delta_{ij}$. Following the approach in \cite{Mitra:2018ezo}, $\sigma_{ij}$ is treated as a perturbative correction to the standard canonical structure $\delta_{ij}$. Since the trace of the identity in three dimensions is $\text{Tr}(\delta_{ij}) = 3$, the trace of the deformation tensor, $\sigma = \delta_{ij}\sigma_{ij}$, must satisfy the constraint $|\sigma| < 1$ to remain within the valid perturbative regime. This ensures that the non-commutative effects represent a consistent deformation of the classical Newtonian framework rather than a complete departure from the underlying symplectic geometry.} Since} $\frac{\Omega_0}{1-\sigma},$ acts as an effective density parameter, it should always be positive. In an ideal fluid structure  Hubble function  deals with two fundamental degrees of freedom $\Omega_0$ and $H_0$. In this modified model the NC correction term  $\sigma$ is also playing the role of an external d.o.f.}

\section{Background Cosmic Dynamics}
\label{BC}
{
 If NC correction term $\sigma$ is set to zero in \eqref{Hubble}, we obtain a Hubble function $$H^2 = H^2_{0}\Bigg[\Omega_{0}(1+z)^{3} + (1- \Omega_{0})(1+z)^{2}\Bigg],$$ that includes contributions from matter and curvature sector. So, it is apparent that $\sigma$ will bring a significant change in the evolution picture.
 We examine the  effect of the inclusion of $\sigma$ on the background cosmic dynamics. }
 
We start the discussion with the deceleration parameter, defined as
\begin{equation}
    q = -1 - \frac{\dot{H}}{H^2}.
\end{equation}
We analyze the impact the  of $\sigma$ on the deceleration parameter. Using \eqref{eqn-acceleration} and \eqref{Hubble}, we reach  the expression of the modified deceleration parameter $q(z)$ shown below
\begin{equation}
    q(z) = \sigma + \frac{1}{2} \frac{\Omega_0 (1+z)^{(3+\sigma)}}{\Bigg[\frac{\Omega_{0}}{1-\sigma}(1+z)^{(3+\sigma)} + (1- \frac{\Omega_{0}}{1-\sigma})(1+z)^{(2\sigma+2)}\Bigg]}.
    \label{decelaration}
\end{equation}
\begin{figure}
    \centering
    \includegraphics[width=0.9\linewidth]{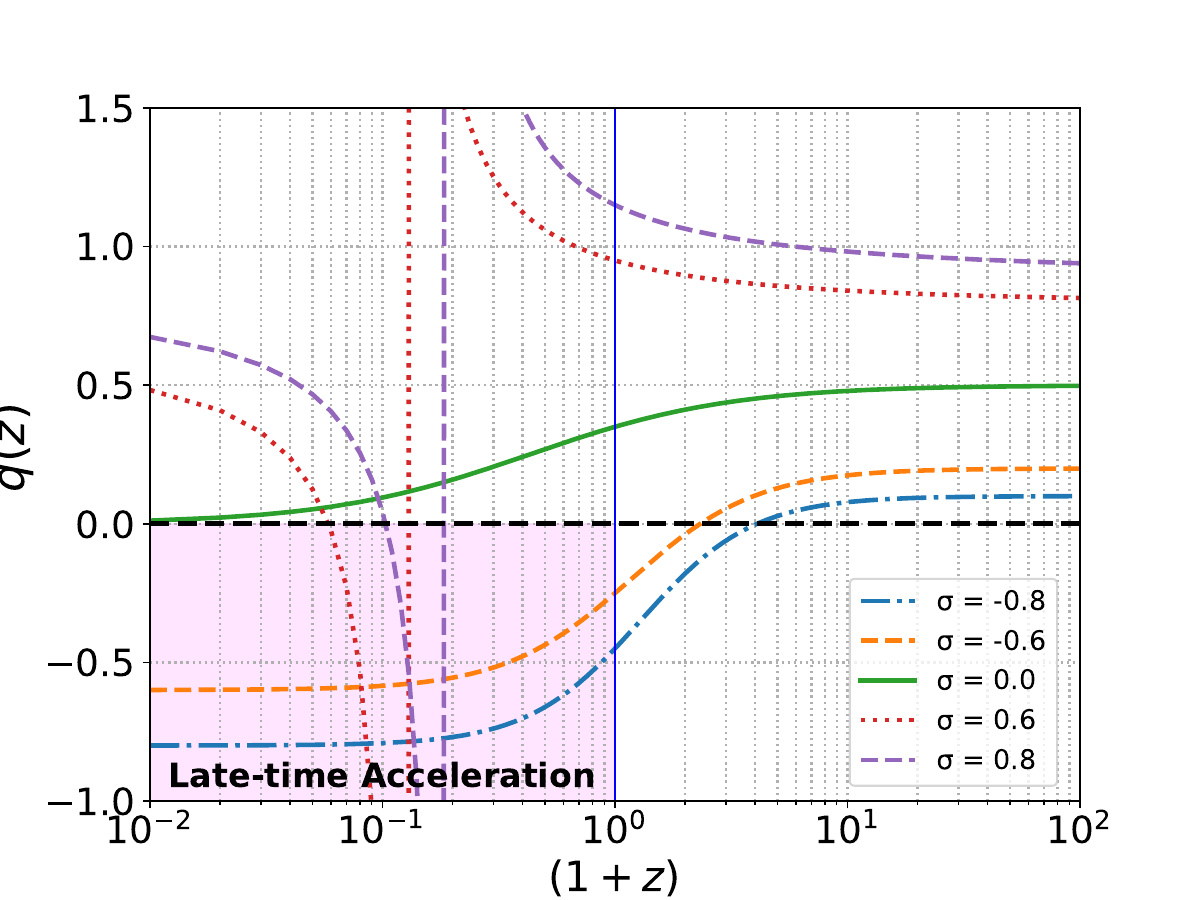}
    \caption{The evolution of deceleration parameter $q(z)$ for different choice of $\sigma$ .}
    \label{fig:enter-theoq}
\end{figure}
\begin{figure}
    \centering
    \includegraphics[width=0.9\linewidth]{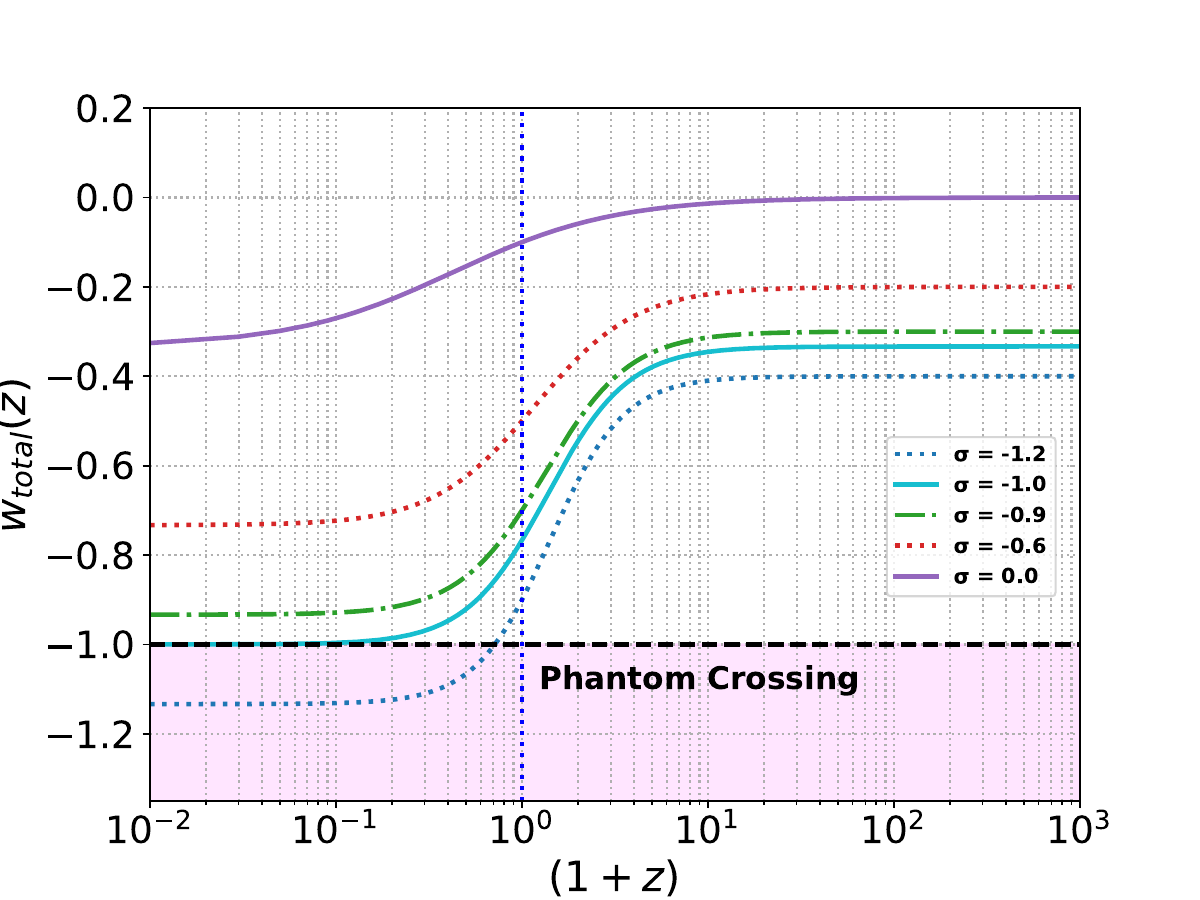}
    \caption{The evolution of overall equation of state parameter for different choice of $\sigma$}
    \label{fig:enter-w}
\end{figure}
    
The above expression describes the evolution of the deceleration parameter $q(z)$, which provides valuable insights into the late-time acceleration of the universe. This behaviour is illustrated in Fig.~\ref{fig:enter-theoq}, where we examine the nature of $q(z)$ for a range of physically admissible $\sigma$ values, as discussed in the previous section, along with a theoretically motivated choice of $\Omega_0$.

We have observed that for $\sigma > 0$, $q(z) > 0$ at very early times (higher $z$ values). We know that $$q(z) = -\frac{\ddot{a}}{H^2 a},$$ we can infer that for $\sigma > 0$, we get $\ddot{a} < 0$, which corresponds to a decelerating phase of cosmic expansion.  Moreover,   in this case, the choice of $\sigma > 0$ results in a discontinuity in $q(z)$ in the $z < 0$ region (i.e., in the future). This intriguing behaviour could be interpreted as a potential 'singularity'. { A more detailed analysis allows us to explore the nature of the singularity. From Equation~\eqref{decelaration}, we observe that if the denominator term,
\begin{equation*}
    \left[ \frac{\Omega_{0}}{1-\sigma}(1+z)^{(3+\sigma)} + \left(1- \frac{\Omega_{0}}{1-\sigma}\right)(1+z)^{(2\sigma+2)} \right],
\end{equation*}
vanishes, the deceleration parameter $q(z)$ will diverge. This defines a critical redshift, $z_s$, characterizing the singularity. By setting the condition:
\begin{equation}
    \left. \left( \frac{\Omega_{0}}{1-\sigma}(1+z)^{(3+\sigma)} + \left(1 - \frac{\Omega_{0}}{1-\sigma}\right)(1+z)^{(2\sigma+2)} \right) \right|_{z_s} = 0,
\end{equation}
and solving for $z_s$, we obtain:
\begin{equation}
    z_{s} = -1 + \left( 1 - \frac{1-\sigma}{\Omega_0} \right)^{\frac{1}{1-\sigma}}.
\end{equation}}

{ We can evaluate the corresponding scale factor at this redshift($a_s= \frac{1}{1+z_s}$). As an example from that plot for $\sigma>0$ values like $\sigma=0.6$ and $0.8$ we will get $z_s = -0.88$ and $-0.81$. So According to \cite{Nojiri:2005sx}, here at this redshift the corresponding Hubble($H_s$)and scale factor $a_s$ and also $\rho_{0,s}$ are finite only $q(z_s)$
diverges. So it is kind of Type II sudden singularity.} For $\sigma > 0$, $q(z)$ does not enter into the late-time accelerating phase, which is inconsistent with the observational evidence.\\
{In the absence of $\sigma$ (i.e., $\sigma = 0$), we also observe a phase of negative acceleration in both the early universe and at present. However, at very future times (i.e., $z < 0$), the acceleration tends to zero, which implies that the universe is approaching a static state. This suggests an  aspect of a non-accelerating universe scenario.\\
The center of interest is the negative-valued region of $\sigma$. In this case, we are achieving cosmic acceleration both at the present day and in the very future.  We have also observed that lower negative values of $\sigma$ cause the entry of $q(z)$ into the zone $q(z) < 0$ at relatively higher redshift. Therefore, it is theoretically clear that in the negative-valued zone of $\sigma$, our model can explain cosmic acceleration.}\\

We have also examined the behaviour of the overall equation of state with respect to our model. We can define our overall equation of state as,
\begin{equation}
    w_{total} = -1  - \frac{2\dot{H}}{3H^2}.
\end{equation}

Using ~\eqref{Hubble}, we provide the expression of $w_{total}$ below.

\begin{widetext}
    \begin{align}
    w_{total}(z) = -1 +  \frac{1}{3}\frac{\Bigg[\frac{(3+\sigma)\Omega_{0}}{1-\sigma}(1+z)^{(3+\sigma)} + (2+2\sigma)(1- \frac{\Omega_{0}}{1-\sigma})(1+z)^{(2\sigma+2)}\Bigg]}{\Bigg[\frac{\Omega_{0}}{1-\sigma}(1+z)^{(3+\sigma)} + (1- \frac{\Omega_{0}}{1-\sigma})(1+z)^{(2\sigma+2)}\Bigg]}
    \label{eqn-eos}
\end{align}
\end{widetext}
{
The expression ~\eqref{eqn-eos} highlights the behaviour of the expansion dynamics of the universe based on our model. The additional term deviating from $-1$ in ~\eqref{eqn-eos} is generated from $\dot{H}$, which essentially represents the change in the expansion rate.
As in the case of $q(z)$, we have also analysed the evolution of the overall equation of state parameter $w_{total}$ for different choices of $\sigma$. In the absence of $\sigma$ (which means $\sigma = 0$), we observed that $w_{total}(z)$ approaches zero on the redshift region $z \geq 10$. In the region of $z < 10$, we can observe that $w_{total}(z)$ trends towards entering the negative-valued zone, and this behaviour is sustained in the future. Thus, in the absence of $\sigma$, we can see a transformation of $w_{total}$ from a dust-like nature ($w_{total} = 0$) to an expanding scenario ($w_{total} < 0$).} 
{
However, when the values of $\sigma$ become negative, we observe the quintessential behaviour of the cosmic expansion dynamics. An interesting fact here is that when $\sigma$ is less than $-1$, we can find a phantom-like behaviour, but this violates the theoretical bound on $\sigma$. Therefore, our model exhibits proper quintessence-like behaviour in the background cosmic dynamics. We have highlighted this fact in Fig-\ref{fig:enter-w}.
  }

\section{Energy Conditions}
Energy conditions are a set of mathematical tools—specifically inequalities—within the framework of general relativity and cosmology. They illuminate the realistic behavior of the energy-momentum tensor in the presence of geometric effects. Furthermore, these conditions characterize the structural properties of cosmological models, providing a physical basis for the dynamics of the universe as predicted by those models. Furthermore, they serve to demonstrate the phase of a model by distinguishing whether it is dominated by the attractive nature of gravity or by a repulsive force associated with negative pressure. We have to first define the fundamental background cosmological equations;
\begin{equation}
    H^2 = \frac{8 \pi G \rho_{eff}}{3},
    \frac{\ddot{a}}{a} = - \frac{4 \pi G}{3}\bigg(\rho_{eff} + 3 P_{eff}\bigg) 
    \label{bc1}
\end{equation}
According to the framework of energy conditions, we establish several fundamental mathematical requirements, each carrying a distinct physical interpretation~\cite{Visser:1997tq,Visser:1999de}:

\begin{itemize}
    \item \textbf{NEC (Null Energy Condition):} The required condition is $\rho_{eff} + P_{eff} \geq 0$. Physically, it ensures that energy density cannot be measured as negative by a null observer and is a prerequisite for the stability of the vacuum.
    
    \item \textbf{WEC (Weak Energy Condition):} This condition requires both $\rho_{eff} \geq 0$ and $\rho_{eff} + P_{eff} \geq 0$. It asserts that the energy density measured by any local observer is non-negative, ensuring the physical validity of the energy source.
    
    \item \textbf{SEC (Strong Energy Condition):} The requirement here is $\rho_{eff} + 3P_{eff} \geq 0$. This condition implies that gravity remains an attractive force. Crucially, the violation of the SEC ($\rho_{eff} + 3P_{eff} < 0$) characterizes a repulsive, anti-gravitational effect, which is the hallmark of "Dark Energy" or accelerated cosmic expansion.
\end{itemize}

\section{Verification of Energy Conditions}

In our model, the effective energy density is expressed as:
\begin{align}
    \rho_{eff} = \frac{3H^2}{8 \pi G} = \nonumber \\ \frac{3H_0^2}{8 \pi G}\Bigg[\frac{\Omega_{0}}{1-\sigma}(1+z)^{3+\sigma} + \left(1- \frac{\Omega_{0}}{1-\sigma}\right)(1+z)^{2\sigma+2}\Bigg].
\end{align}

For a negative non-commutative parameter ($\sigma < 0$), the coefficients $\frac{\Omega_{0}}{1-\sigma}$ and $1 - \frac{\Omega_{0}}{1-\sigma}$ remain positive (given that $\Omega_0 < 1$). Within the theoretical bounds of $-1 < \sigma < 1$, it is evident that the reduced Hubble parameter $E^2 = H^2/H_0^2$ remains positive throughout the cosmic evolution. Consequently, we confirm that the condition \textbf{$\rho_{eff} \geq 0$} holds, ensuring the physical validity of the energy density despite the negative values of $\sigma$.

To evaluate the Null Energy Condition (NEC), defined as $\rho_{eff} + P_{eff} \geq 0$, we consider the continuity equation for the effective fluid:
\begin{equation}
    \dot{\rho}_{eff} + 3H(\rho_{eff} + P_{eff}) = 0.
\end{equation}

Using the standard relations $\dot{a} = aH$ and $a = 1/(1+z)$, the time derivative of the energy density can be transformed into a derivative with respect to redshift:
\begin{equation}
    \dot{\rho}_{eff} = \frac{d\rho_{eff}}{dz} \frac{dz}{dt} = -H(1+z) \frac{d\rho_{eff}}{dz}.
\end{equation}

Substituting this into the continuity equation, we obtain:
\begin{equation}
    \rho_{eff} + P_{eff} = \frac{(1+z)}{3} \frac{d\rho_{eff}}{dz}.
\end{equation}

Therefore, for the NEC to be satisfied ($\rho_{eff} + P_{eff} \geq 0$), the following inequality must hold:
\begin{equation}
    (1+z)\frac{d\rho_{eff}}{dz} \geq 0.
\end{equation}

Applying this to our specific model, we derive:
\begin{align}
    (1+z)\frac{d\rho_{eff}}{dz} = \frac{3H_0^2}{8 \pi G} \Bigg[ \frac{\Omega_{0}(3+\sigma)}{1-\sigma}(1+z)^{3+\sigma} + \nonumber \\ (2+2\sigma) \left(1- \frac{\Omega_{0}}{1-\sigma}\right)(1+z)^{2\sigma+2} \Bigg].
\end{align}

Based on the same logic applied to $\rho_{eff} \geq 0$, it is clear that for $-1 < \sigma < 1$, every term in the bracket remains positive. Thus, $(1+z)\frac{d\rho_{eff}}{dz} \geq 0$ is consistently satisfied. We can therefore conclude that both the \textbf{NEC} and \textbf{WEC} are strictly satisfied within the theoretical bounds of the model.

Finally, we examine the Strong Energy Condition (SEC). To evaluate this, we refer to the modified acceleration equation \eqref{eqn-acceleration_new}:

\begin{equation}
\frac{\ddot{a}}{a} = -\frac{4\pi G}{3} \left( \rho_{0} + \frac{3\sigma H^2}{4 \pi G} \right).
\label{eqn-acceleration_new}
\end{equation}

In accordance with the standard definition of the SEC ($\rho_{eff} + 3P_{eff} \geq 0$), the gravitational acceleration remains attractive only if the following inequality holds:
\begin{equation}
    \rho_{0} + \frac{3\sigma H^2}{4 \pi G} \geq 0.
\end{equation}

Since the Hubble parameter $H^2$, the matter density $\rho_0$, and the gravitational constant $G$ are all positive quantities, this condition can only be satisfied if $\sigma \geq 0$. However, a positive or zero value for $\sigma$ would result in a decelerating universe ($\ddot{a} \leq 0$), which contradicts current cosmological observations.

This result confirms that the SEC must be violated to account for the observed late-time acceleration. For negative values of the non-commutative parameter ($\sigma < 0$), the term $\frac{3\sigma H^2}{4\pi G}$ acts as a source of repulsive gravity, or "anti-gravity," effectively driving the cosmic acceleration. 

In summary, we have consistently demonstrated that for the parameter range $\sigma < 0$, both the \textbf{NEC} and \textbf{WEC} are satisfied, ensuring the physical stability of the model, while the \textbf{SEC} is violated. Consequently, this model successfully provides a theoretical basis for the accelerated expansion of the universe through the consideration of non-commutative geometric effects.
\section{Observational Dataset and Methodology}
\label{data}
In this section, we will mention the observational datasets that have been used for our work. We would also discuss the statistical methods for performing observational formalities. 

Our model is characterised by two fundamental degrees of freedom, $H_0$ and $\Omega_0$, along with  parameter $\sigma$, generated from non-commutative (NC) corrections, which serves as the primary theoretical input. The parameter $\sigma$ is of particular interest in this study, and it is essential to incorporate observational constraints to better understand its physical relevance.


The following datasets have been used for analysis.

\begin{itemize}
    \item \textbf{OHD:} One of the most fundamental and primary datasets used to constrain cosmological models is the $H(z)$ dataset. In this work, we utilise an updated collection of 57 data points of $H(z)$, which we refer to as OHD, as presented in~\cite{Bouali:2023flv}. Out of these 57 data points, 31 are derived from Cosmic Chronometer (CC) measurements, which involve the differential age (DA) between evolving galaxies, while the remaining 26 data points are obtained from Baryonic Acoustic Oscillations (BAO) measurements related to galaxy clustering phenomena.
    \item \textbf{SNeIa:} We have used the following type Ia supernovae datasets for our analysis.
    \begin{itemize}
    \item [1.]\textbf{Union3:} The latest Union3 compilation includes 2087 SNeIa from 24 datasets, standardized to a consistent distance scale via SALT3 light curve fitting. These are analysed and binned using the UNITY1.5 Bayesian framework, yielding 22 binned distance modulus measurements in the redshift range $0.05 < z < 2.26$, which we adopt in this article~\cite{Rubin:2023ovl}.

    \item[2.] \textbf{Pantheon+:} This type Ia Supernovae (SNeIa) data is an updated version of the  Pantheon sample. It includes 1701 light curve measurements from 1550 distinct supernovae. This sample is activated within the redshift range $z = [0.001, 2.26]$ ~\cite{Brout:2022vxf}. We have implemented here 1590 data points from this sample to demolish the effect of peculiar velocity issues ~\cite{Chan-GyungPark:2024mlx}. { The influence of the 'SH0ES'(Supernova H0 for the Equation of
     State of Dark energy) calibration is inherently present in our analysis.}
    {\item[3.]\textbf{DESY5:}We use a five-year dataset from the Dark Energy Survey (DES) Supernova Program, which includes 1635 SNeIa in the redshift range $0.1 < z < 1.3$, along with 194 low-redshift SNeIa from CfA3, CfA4, CSP, and Foundation surveys ($0.025 < z < 0.1$) ~\cite{DES:2024jxu}}.
    \end{itemize}
    \item \textbf{DESI :} The Dark Energy Spectroscopic Instrument (DESI) data includes both isotropic and anisotropic Baryonic Acoustic Oscillations (BAO) measurements, which are based on BAO matter clustering as a standard ruler. For isotropic BAO measurements, the angle-averaged distance rescaled by the co-moving sound horizon ($r_d$) at the drag epoch, $D_V(z)/r_d$, plays a key role. In contrast, anisotropic BAO measurements are represented by $D_M(z)/r_d$ and $D_H(z)/r_d$, where $D_M$ denotes the co-moving distance and $D_H$ refers to the Hubble horizon distance. For this study, we have utilized data from DESI Data Release I (\textbf{DESI DR1}), which includes tracers from the Bright Galaxy Sample (BGS), Luminous Red Galaxies (LRG), Emission Line Galaxies (ELG), Quasars (QSO), and the Lyman-$\alpha$ forest. The complete DESI dataset used in this work is presented in Table I of ~ \cite{DESI:2024mwx}. We also have utilised the recently updated DESI Data Release II(\textbf{DESI DR2}), which is collected from ~\cite {DESI:2025zgx}. 
\end{itemize}

We have utilized the Markov Chain Monte Carlo (MCMC) method to constrain our model using the observational datasets mentioned above. For this purpose, we have utilized the publicly available Python-based package \textbf{\texttt{emcee}}\footnote{\href{https://github.com/dfm/emcee}{https://github.com/dfm/emcee}} ~ \cite{Foreman-Mackey:2012any}. The \textbf{\texttt{emcee}} sampler efficiently explores the posterior distribution in high-dimensional spaces by using an ensemble of walkers. We marginalize the likelihood across the datasets with priors for the free parameters, as specified in Table~\ref{tab:priors}. Additionally, we used the \textbf{\texttt{getdist}} package\footnote{\href{https://github.com/cmbant/getdist}{https://github.com/cmbant/getdist}} ~\cite{Lewis:2019xzd} to analyze our MCMC chains, ensuring the proper physical interpretation of our cosmological parameters~\cite{Lewis:2019xzd}.

\begin{table}
    \centering
    \begin{tabular}{|c|c|}
    \hline
        ~~~~~~Parameters~~~~~~    &  ~~~~~~Priors~~~~~~  \\
    \hline 
    \hline
         ~~~~~~$H_{0}$~~~~~~      &   ~~~~~~[50, 100]~~~~~~       \\
         ~~~~~~$\Omega_{0}$~~~~~~ &   ~~~~~~[0, 1]~~~~~~        \\
         ~~~~~~$\sigma$~~~~~~     &   ~~~~~~[-2.9, 0.9]~~~~~~        \\
        
    \hline 
         
    \end{tabular}
    \caption{Priors on cosmological parameters are constrained using the observational datasets discussed.
}
    \label{tab:priors}
\end{table}

\section{Results}
\label{RE}
In this section, we will provide a comprehensive discussion of the overall results from our analysis, focusing on key findings and their implications. We have examined the observational constraints within 1-$\sigma$ uncertainties, and the results for different data sets are presented in Table-\ref{tab-mcmc}. These constraints are crucial for evaluating the accuracy and reliability of our model, as they reflect the range of values that are statistically consistent with the observed data.\\

To start with, we will discuss the observational constraints related to the outcome of the Hubble Data Mission (OHD).
Looking at $H_0$, we observe that the value obtained from the Hubble Data Mission (OHD) is lower. { {We found that the Hubble constant yields a value of $H_0 = 59.7^{+1.1}_{-0.98}$ km s$^{-1}$ Mpc$^{-1}$, which is significantly lower than the standard consensus and thus not a physically suitable estimate for $H_0$. This discrepancy arises because the Cosmic Chronometer (OHD) data alone is insufficient to  provide robust constraints on the model parameters. Consequently, the parameters are poorly constrained when using the OHD dataset in isolation. For this reason, we have omitted the corresponding 2D marginalized posterior contours, as they do not provide a statistically meaningful representation of the parameter space. }} From the OHD data alone, we obtained $\Omega_0 = 0.80^{+0.28}_{-0.45}$. The model parameter $\sigma$ has been constrained in the negative-valued region ($\sigma = -0.42^{+0.16}_{-0.20}$), which is consistent with our earlier prediction.
 Therefore, we now need to take additional steps to interpret our model more effectively.


Before implementing further missions in our MCMC analysis, we also show the evolution of the distance modulus $\mu(z)$ for our model, along with a comparison to $\Lambda$CDM, using the latest Pantheon+ dataset.
The distance modulus is defined as,
\begin{equation}
\mu(z)=5\log_{10} \frac{{d_{\rm{L}}(z)}}{\rm{Mpc}}+25 ,
\end{equation}%
where the luminosity distance $d_{\rm{L}}(z)$ is
\begin{equation}
d_{\rm{L}}(z)=(1+z)\int_{0}^{z}\frac{dz^{^{\prime }}}{H(z^{^{\prime }})}.
\end{equation}

We are now using the type Ia supernovae (SNeIa) missions data along with OHD. However, with the inclusion of Pantheon+, we observed a significant improvement in the $H_0$ measurements. For the combined OHD+Pantheon+ data, we obtained $H_0 = 73.03 \pm 0.19$, which is very close to the local measurements of $H_0$ ~\cite{Riess:2021jrx}. With OHD+DESI+Pantheon+ we have obtained $H_0 = 72.27 \pm 0.17$. { {While these results approach the values reported by local measurements, caution is required before drawing conclusions regarding the "Hubble Tension." This discrepancy in $H_0$ between local distance ladder observations and CMB-based measurements constitutes the core of this foundational tension. In our current analysis, we have not incorporated CMB data, such as the temperature anisotropy power spectra ($C_\ell^{TT}$ or $D_\ell^{TT}$), and thus any definitive commentary on the resolution of the tension would be premature. 
Furthermore, our analysis of the Pantheon+ sample utilizes the distance modulus rather than the raw apparent magnitude. This approach inherently incorporates the 'SH0ES' calibration, which tends to shift $H_0$ toward higher values due to presence of "Cepheid Host distances" ~\cite{Riess:2021jrx}. Consequently, the observed tension in $H_0$ when comparing Pantheon+ to other datasets is, in part, a reflection of this calibration baseline. In ~\cite{Pan:2025qwy}(in first version), a similar effect is observable in their analysis. }}

For the alternative combination of OHD and the Union3 supernova sample, the constrained value of $H_0$ is notably lower even than the estimates provided by the Planck mission ~\cite{Planck:2018vyg}. It is important to acknowledge that a complete version of the Union3 dataset was not fully available at this time ; instead, we utilized the 22-set binned data. Consequently, the incorporation of the full Union3 sample in future studies may significantly improve the constraints on $H_0$ and provide a more robust statistical baseline. 

 Since $\Omega_0$ is a cosmic parameter, the rescaled quantity $\frac{\Omega_0}{(1-\sigma)}$ acts like an effective density parameter.

The addition of DESI, along with the corresponding SNeIa, causes the $\Omega_0$ values to decrease slightly. It is also observed that DESI DR2 makes the value of $\Omega_0$ a little bit lower in comparison to DESI DR1. 

\begin{figure*}
    \centering
    \includegraphics[width= 0.49\linewidth]{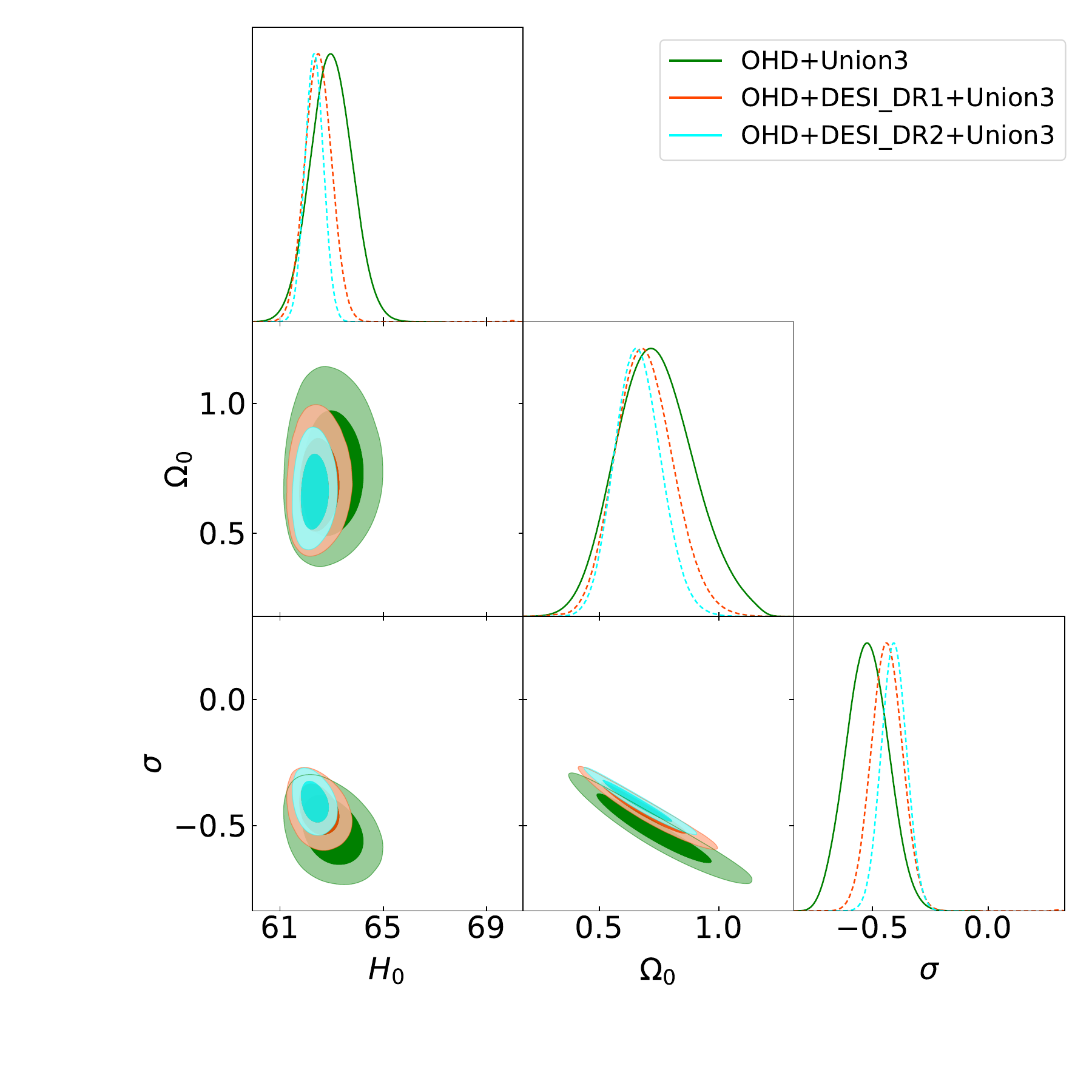}
    \includegraphics[width=0.49\linewidth]{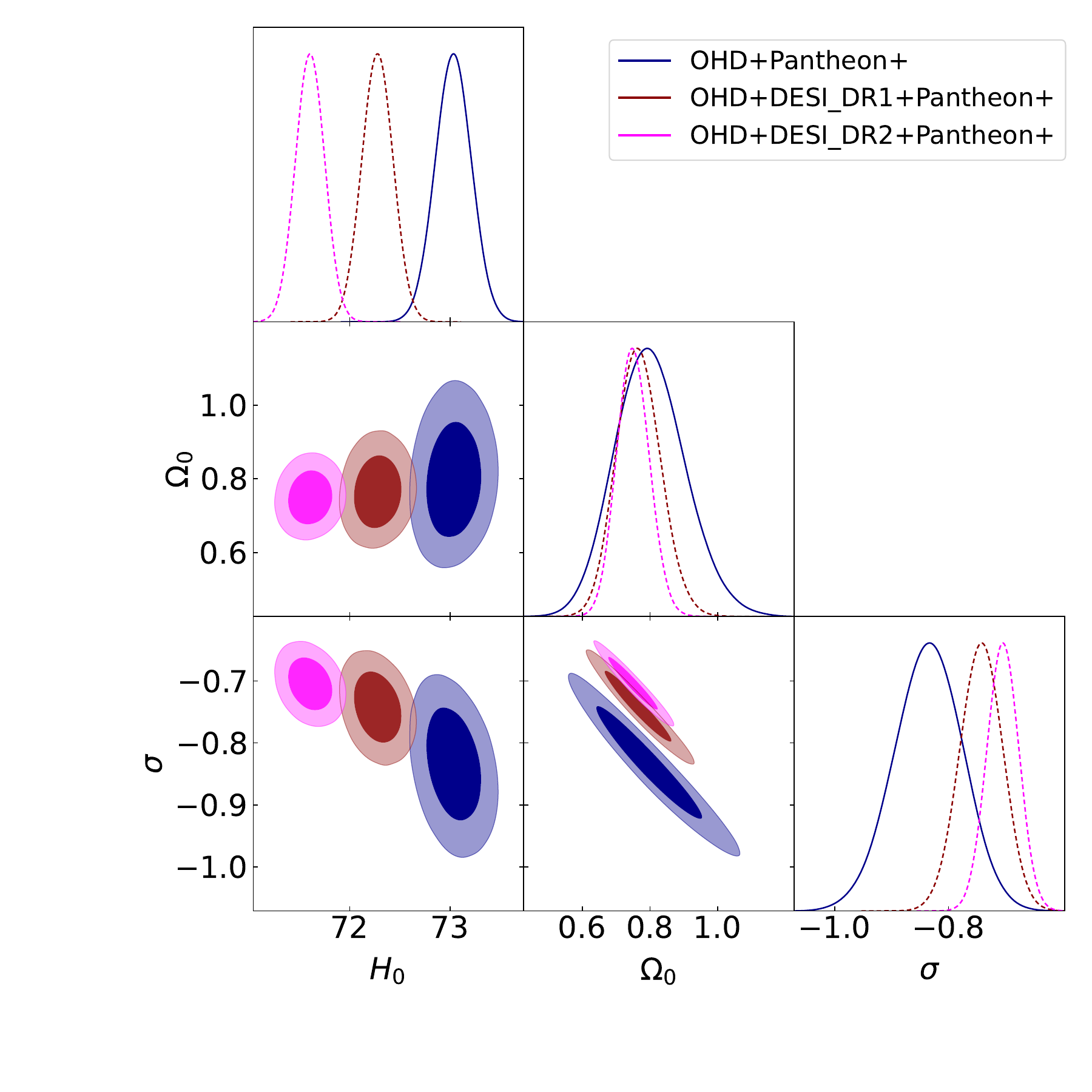}

    \includegraphics[width=0.49\linewidth]{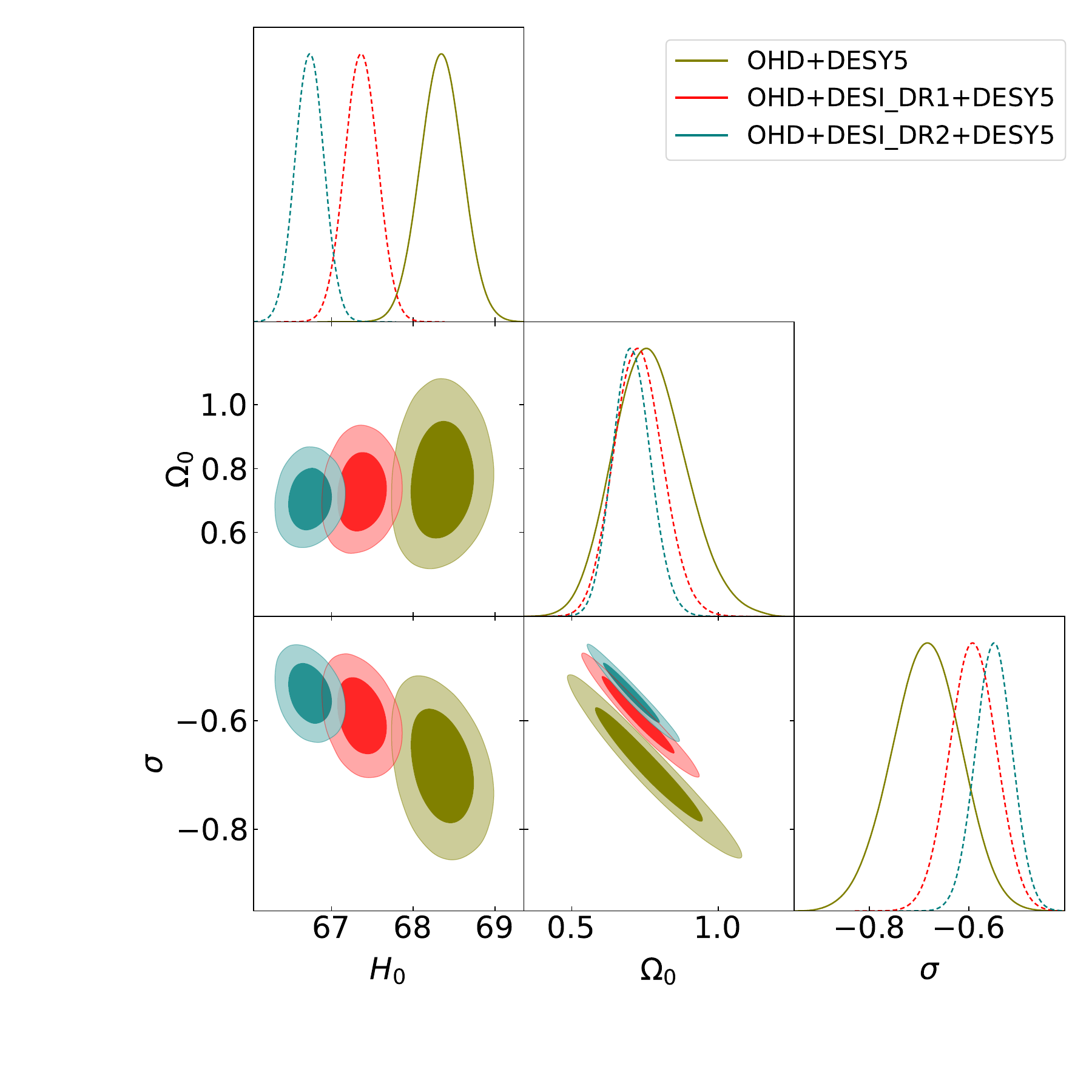}
    \caption{2D contour plots with 1D posterior distributions for different combinations of OHD, DESI, and supernova datasets.}
    \label{2Dcontour}
\end{figure*}

Compared to the other supernova datasets, Pantheon+ improves the 1-$\sigma$ uncertainties. The value of $\sigma$ for Pantheon+ is lower than in the case of Union3. The same trend is observed for $H_0$. This fact is also highlighted in the 2D contour presentation in Fig-\ref{2Dcontour}. $\sigma$ is negatively correlated with both $H_0$ and $\Omega_0$, but the negative correlation is stronger in the $\sigma-\Omega_0$ plane. The addition of DESI(DR1 and DR2) with two different OHD+SNeIa missions also maintains the same correlation properties. DESI has improved the 1-$\sigma$ uncertainties, and the 2D contour area is smaller in the case of the OHD+DESI+SNeIa missions. This reflects the strong constraining power of DESI. When comparing DESI DR1 and DESI DR2, we observe that the contour plots for DESI DR2 appear more refined and well-developed. This indicates that DESI DR2 offers relatively stronger constraining power on the cosmological parameters, likely due to improved data quality or enhanced observational precision.

\begin{figure*}
    \centering
\includegraphics[width=0.7\linewidth]{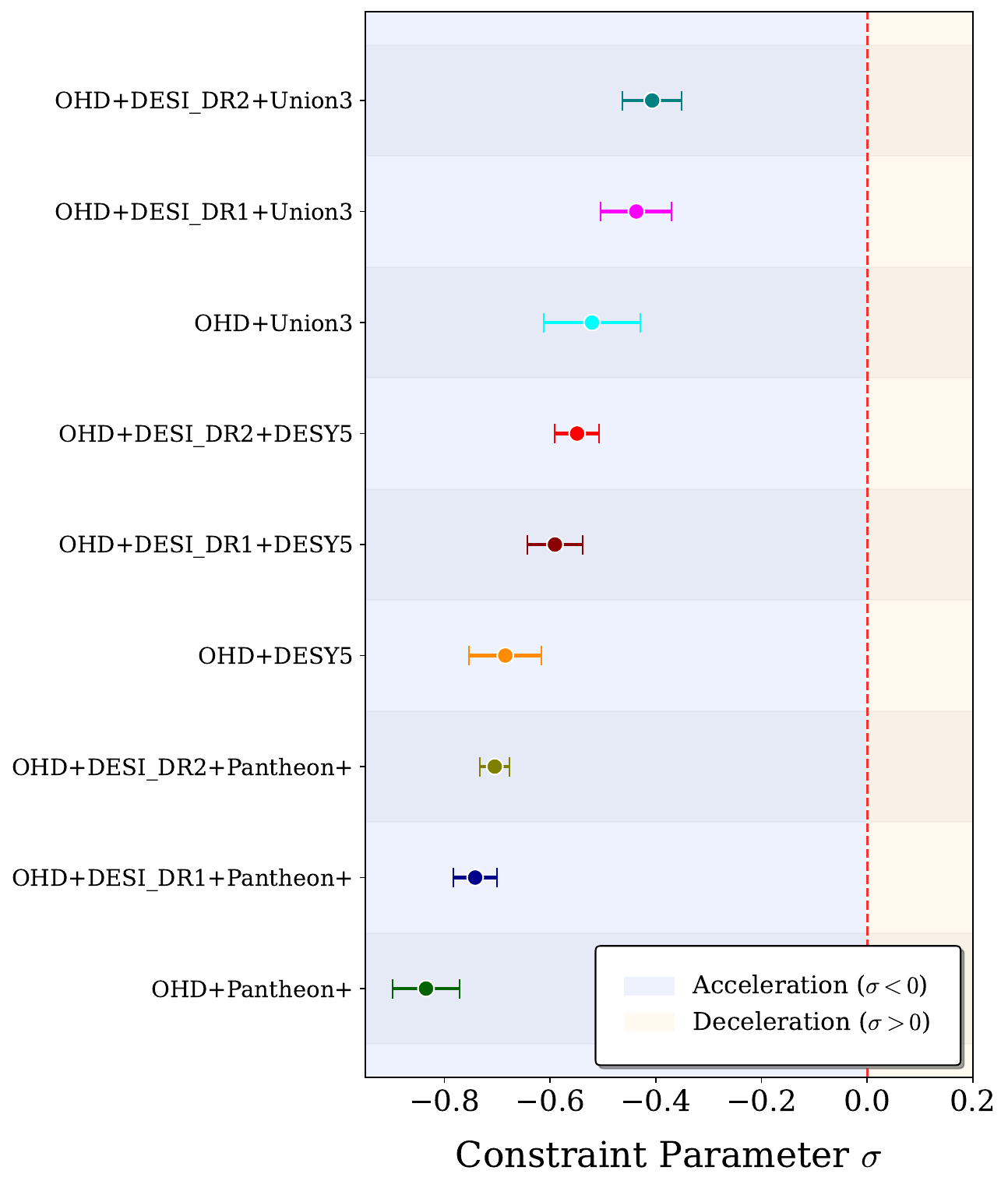}
    \caption{Whisker presentation based on observational constraints on $\sigma$}
    \label{fig:whisker}
\end{figure*}













\begin{table*}[ht]
\centering
\renewcommand{\arraystretch}{1.3} 

\begin{tabular}{lccc}
\toprule
\hline
\textbf{Dataset} & \boldmath{$H_0$} & \boldmath{$\Omega_0$} & \boldmath{$\sigma$} \\
\hline
\hline
\midrule
OHD & $59.7^{+1.1}_{-0.98}$ & $0.80^{+0.28}_{-0.45}$ & $-0.42^{+0.16}_{-0.20}$ \\
\hline
OHD + Union3 & $63.00 \pm 0.81$ & $0.74^{+0.15}_{-0.17}$ & $-0.521 \pm 0.091$ \\

OHD + Pantheon+ & $73.03 \pm 0.19$ & $0.801^{+0.096}_{-0.11}$ & $-0.835 \pm 0.063$ \\

OHD + DESY5 & $68.34 \pm 0.26$ & $0.77^{+0.11}_{-0.13}$ & $-0.685 \pm 0.069$ \\
\hline
\midrule

OHD + DESI DR1 + Union3 & $62.48 \pm 0.54$ & $0.69^{+0.11}_{-0.13}$ & $-0.437 \pm 0.067$ \\

OHD + DESI DR1 + Pantheon+ & $72.27 \pm 0.17$ & $0.766 \pm 0.066$ & $-0.742 \pm 0.041$ \\

OHD + DESI DR1 + DESY5 & $67.36 \pm 0.22$ & $0.729 \pm 0.083$ & $-0.591 \pm 0.052$ \\
\hline
\midrule

OHD + DESI DR2 + Union3 & $62.33 \pm 0.43$ & $0.664^{+0.090}_{-0.10}$ & $-0.407 \pm 0.056$ \\

OHD + DESI DR2 + Pantheon+ & $71.61 \pm 0.14$ & $0.750 \pm 0.048$ & $-0.705 \pm 0.028$ \\

OHD + DESI DR2 + DESY5 & $66.74 \pm 0.24$ & $0.705 \pm 0.066$ & $-0.549 \pm 0.042$ \\
\bottomrule
\hline
\hline
\end{tabular}
\caption{Details of observational constraints within 1-$\sigma$ uncertainties.}
\label{tab-mcmc}
\end{table*}

\begin{figure*}
    \centering
    \includegraphics[width= 0.45\linewidth]{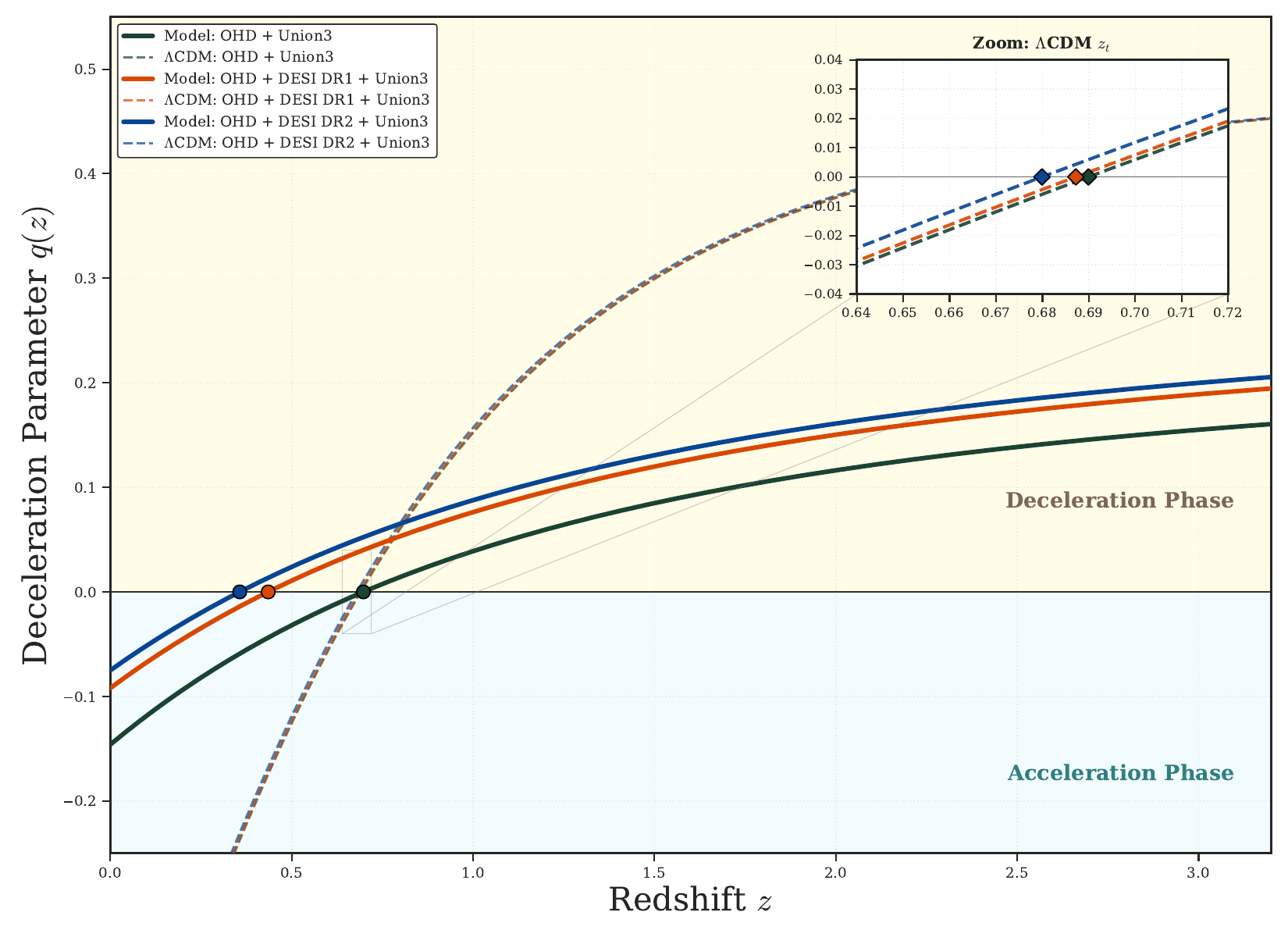}
    \includegraphics[width=0.45\linewidth]{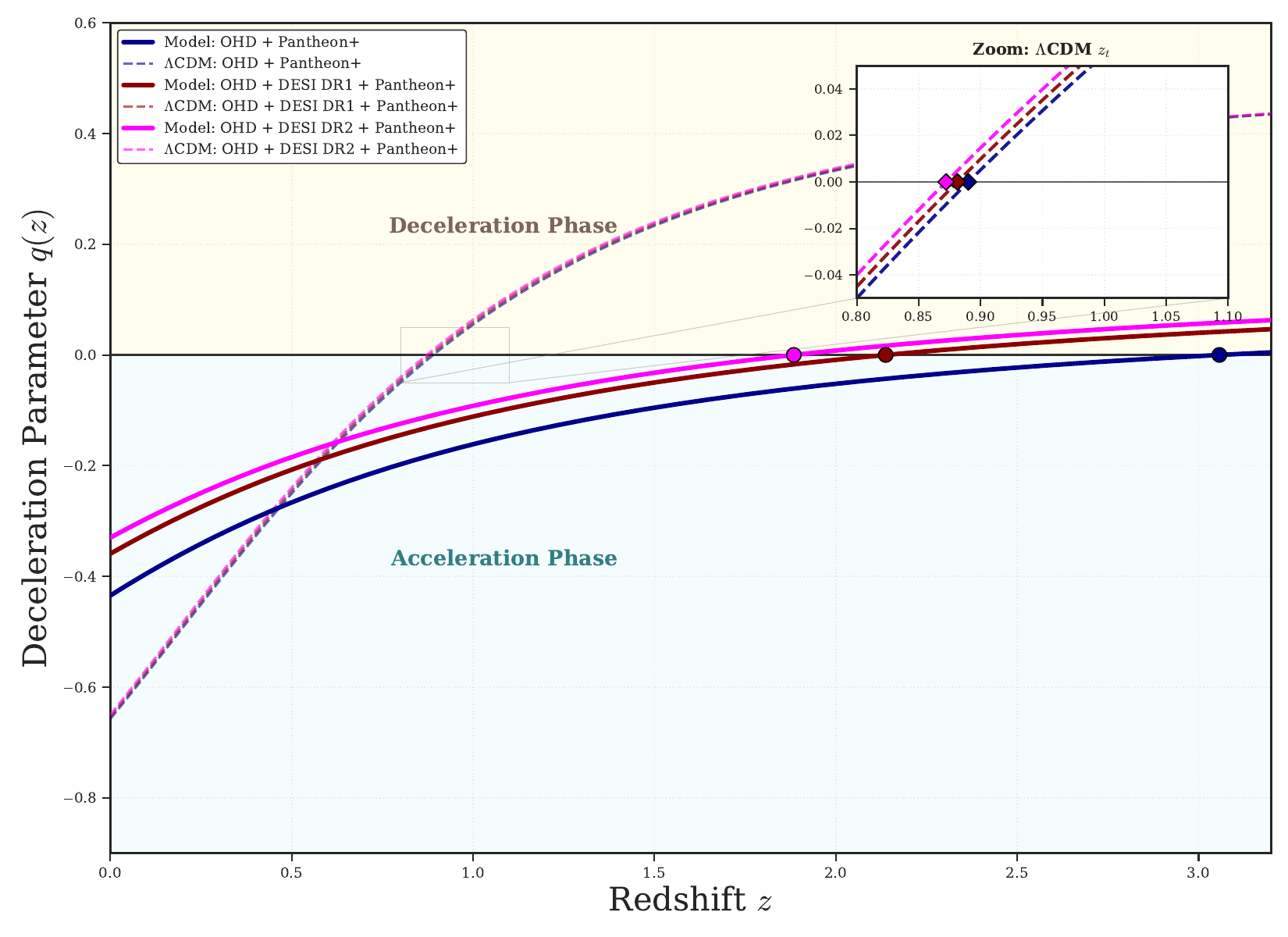}

    \includegraphics[width=0.48\linewidth]{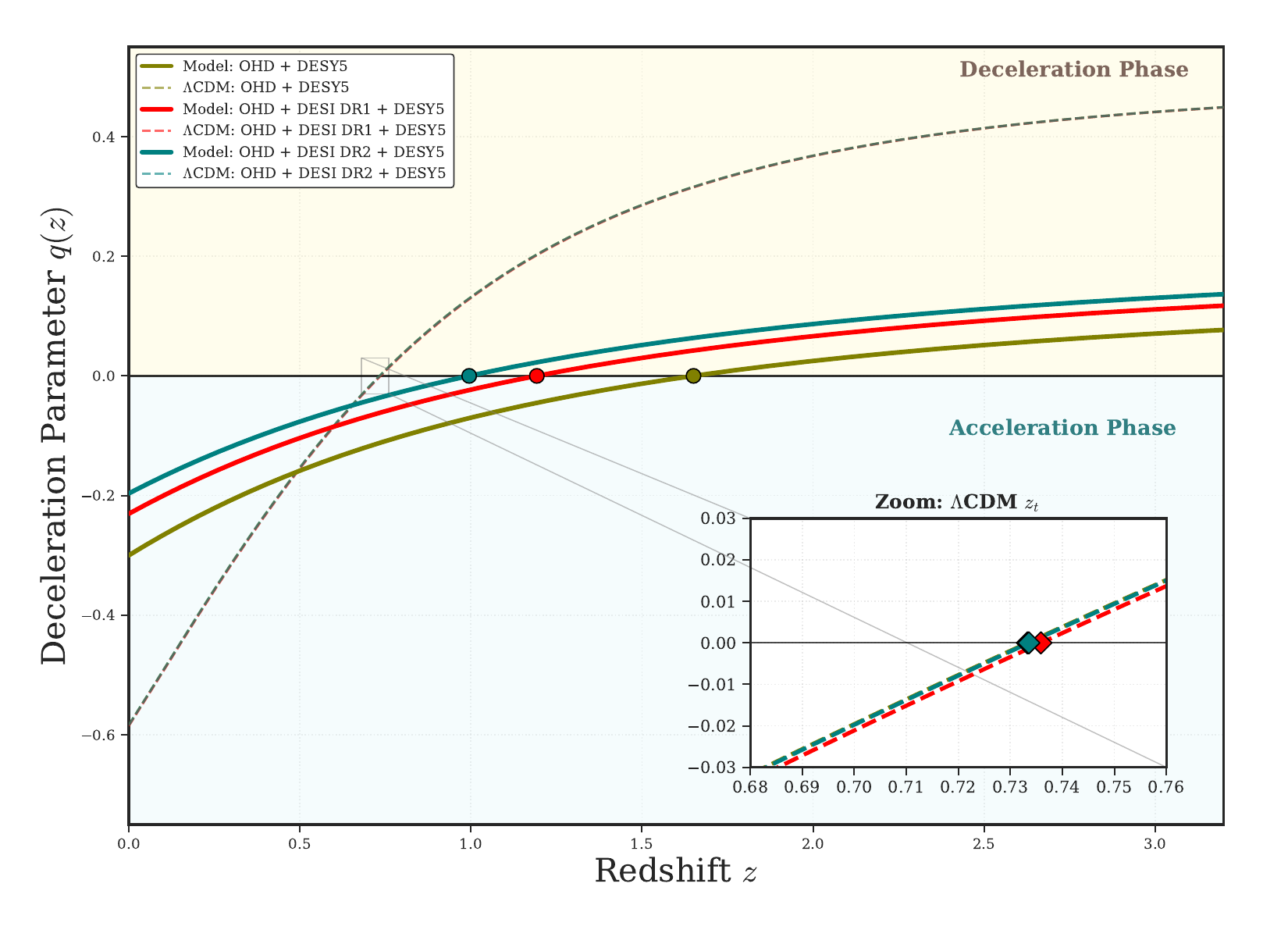}
    \caption{ {The evolution of $q(z)$ on the basis of observational constraints of OHD + SNeIa and OHD + DESI + SNeIa datasets with comparison to $\Lambda$CDM observations}}
    \label{q-obs}
\end{figure*}

We have discussed cosmic acceleration several times, focusing on the observational perspective. Specifically, we examined the nature of acceleration through the evolution of the deceleration parameter, based on observational constraints. This is presented in Fig. \ref{q-obs}. We have observed that the addition of DESI, along with the relative SNeIa, leads to a decrease in the present-day acceleration. The most notable observation is that the present cosmic acceleration appears to be slightly reduced in the case of DESI DR2 compared to DR1 across all SNeTa combinations. DESI not only indicates a slightly reduced present-time acceleration but also suggests that the onset of the present-day acceleration phase occurs somewhat later compared to the OHD+SNeIa combinations. Figure~\ref{q-obs} actually shows this effect. In the case of the Pantheon+ dataset, it is observed that the onset of the accelerating phase occurs slightly earlier compared to Union3. Meanwhile, DESY5 indicates the beginning of acceleration at an intermediate epoch, falling between those inferred from Union3 and Pantheon+. { In addition, we include a comparison with the $\Lambda$CDM model in Fig.~\ref{q-obs}. We observe a transition redshift pattern similar to that of our model, specifically $z_{t,\text{OHD+DESI DR2 + SNeIa}} < z_{t,\text{OHD+DESI DR1 + SNeIa}} < z_{t,\text{OHD+SNeIa}}$ for the $\Lambda$CDM framework. Since the constrained values of $\Omega_{m}$ for $\Lambda$CDM are very close across all observational datasets, it is challenging to distinguish the transition redshifts for each mission individually. To address this, we have provided a zoomed-in inset in Fig.~\ref{q-obs} to resolve these differences. A notable exception is the OHD+DESY5 case, where the $q(z)$ evolution nearly overlaps with the OHD+DESI DR2 + DESY5 dataset. Finally, we note that the $\Lambda$CDM model produces a greater degree of late-time acceleration compared to our proposed model.}

\begin{figure*}
\begin{center}
\begin{subfigure}{}
    \includegraphics[width=0.3\linewidth]{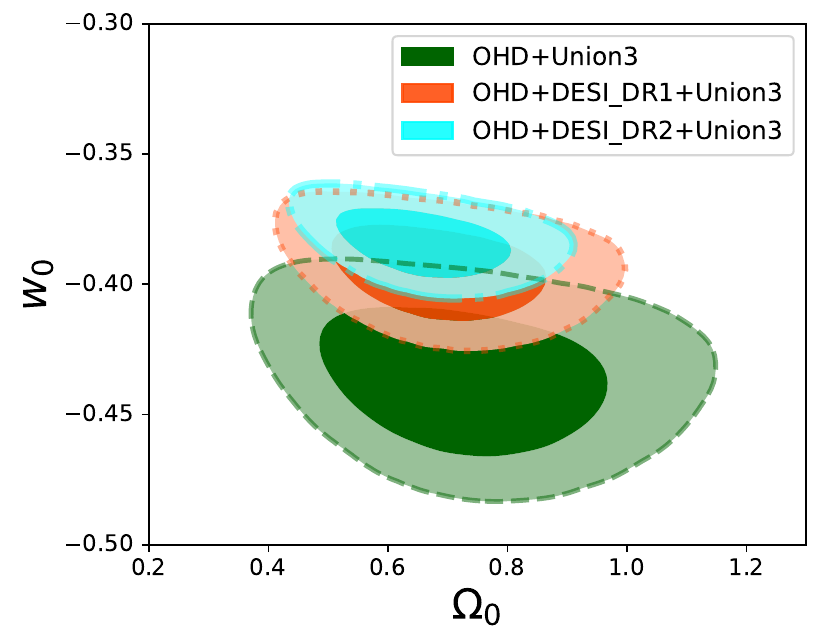}
    \includegraphics[width=0.3\linewidth]{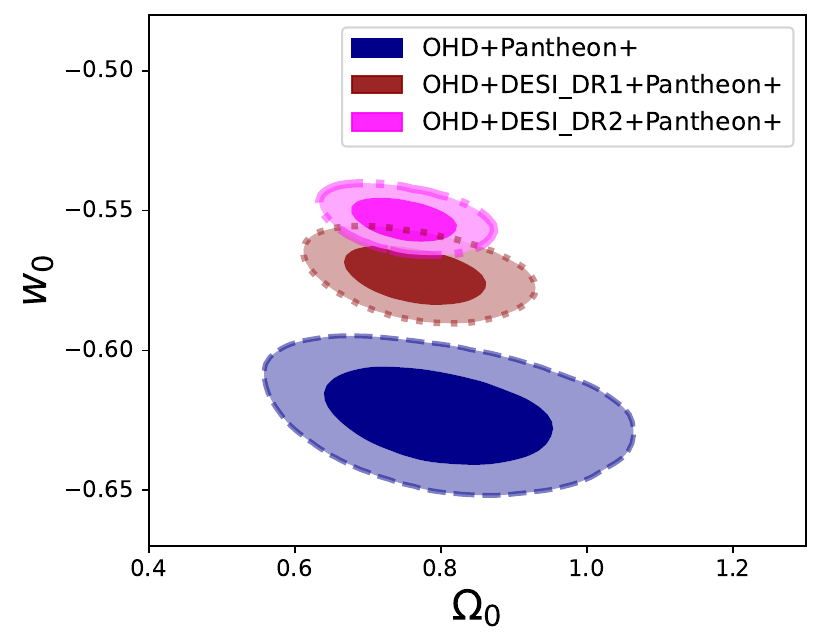}
    \includegraphics[width=0.3\linewidth]{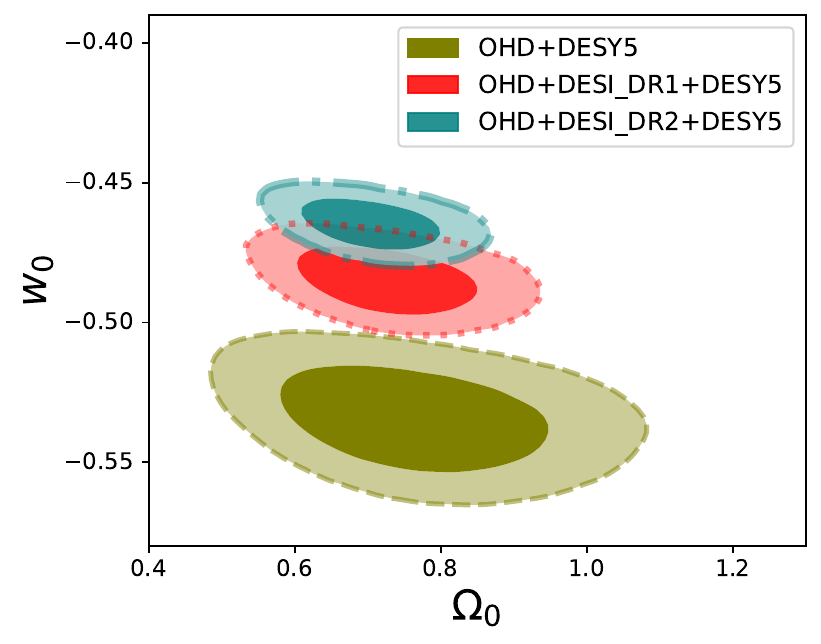}
    
\end{subfigure}
    \begin{subfigure}{}
        
        \includegraphics[width=0.3\linewidth]{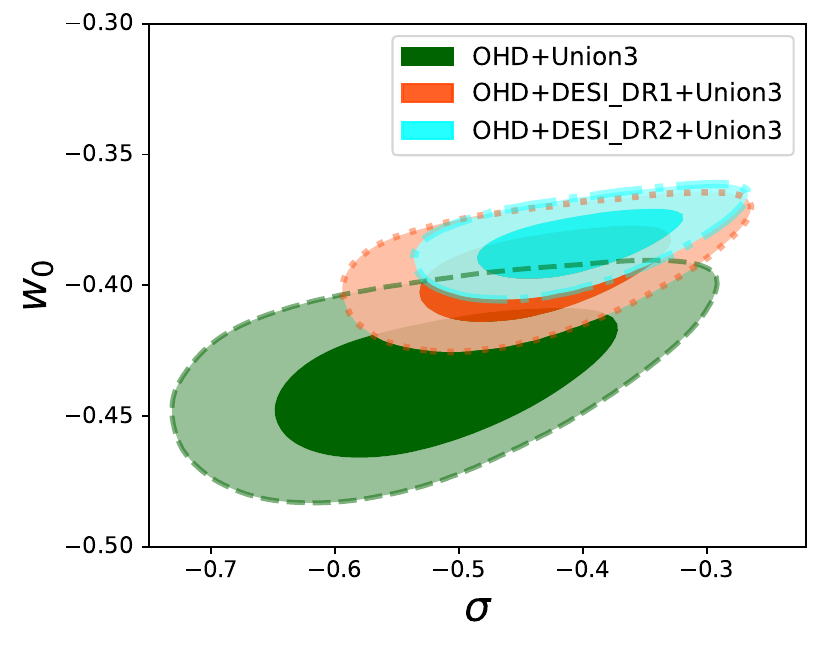}
        \includegraphics[width=0.3\linewidth]{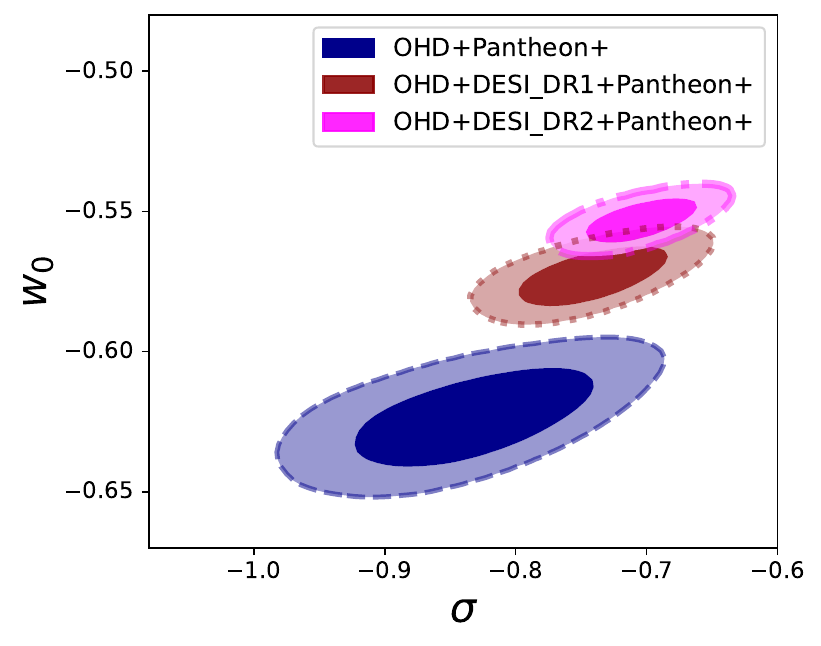}
        \includegraphics[width=0.3\linewidth]{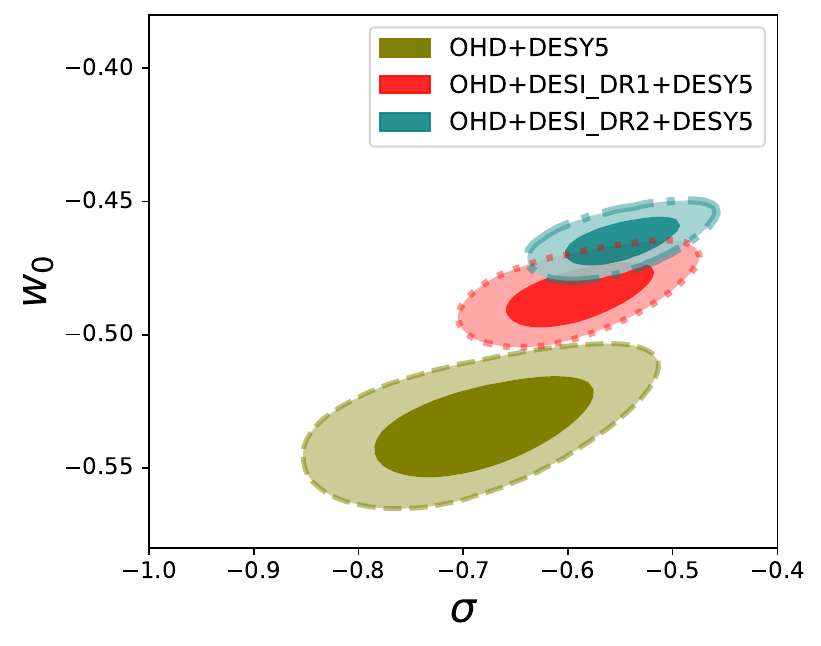}
    \end{subfigure}
    
\end{center}
\caption{ The derived 2D contour of the present-day overall equation of state parameter based on the observational constraints from OHD + SNeIa and OHD + DESI + SNeIa missions.}
\label{fig:derived}
\end{figure*}

We have also derived the 2D contour for the overall equation of state $w_{total}$ based on the observational constraints of OHD + SNeIa, OHD + DESI DR1 + SNeIa, and OHD +  DESI DR2 + SNeIa missions. This is presented in Fig-\ref{fig:derived}. The 2D contour of $w_{0}-\Omega_0$ and $w_{0}-\sigma$ makes it clear that our model exhibits quintessence-like behaviour in the context of cosmic acceleration. In all the cases, $w_0$ is negatively correlated with $\Omega_0$, and it has a positive correlation with the model parameter $\sigma$.

The parameter $\sigma$ represents the primary contribution arising from the noncommutative (NC) correction. In this work, we have gone through a detailed analysis to investigate the observational bound on $\sigma$. This feature is further illustrated through the whisker plot shown in Fig~\ref{fig:whisker}. From this visual representation, it becomes evident that the parameter $\sigma$ remains consistently less than zero across all dataset combinations, providing clear support for a negative $\sigma$ in each case. Although in the case of OHD + DESY5, we observe that the error bar touches the $\sigma=0$ boundary at the 1$\sigma$ confidence level, the inclusion of DESI data helps to recover the $\sigma < 0$ scenario even in the DESY5 case.

\section{Goodness of fit and Model Comparison }
\label{MC}
In this work, we have derived a generalized cosmological model based on Newtonian cosmology incorporated with a non-commutative term. Having established the fundamental cosmological framework of this model, it is essential to assess its observational viability and statistical robustness. To evaluate these properties, we perform a $\chi^2_{\text{min}}$ (minimum chi-squared) analysis to determine the best-fit parameters. Furthermore, we provide a comparative analysis of our results against the well-established vanilla $\Lambda$CDM model to benchmark the statistical quality of our proposed model. For the purpose of our analysis, the $\chi^2$ functions are defined as follows:
 \begin{itemize}
    \item \textbf{OHD:} 
    \[ \chi^2_{\text{OHD}} = \sum_{i=1}^{N} \left( \frac{H_i - H(z_i, \theta)}{\sigma_i} \right)^2 \]

    \item \textbf{BAO:} 
    \[ \chi^2_{\text{BAO}} = \sum_{i,j} (\mu_i - \mu_{Th}(z_i, \theta))^T C_{\text{BAO},ij}^{-1} (\mu_j - \mu_{Th}(z_j, \theta)) \]

    \item \textbf{SNeIa:}
    \[ \chi^2_{\text{SNeIa}} = \sum_{i,j} (\mu_i - \mu_{Th}(z_i, \theta))^T C_{\text{SNeIa},ij}^{-1} (\mu_j - \mu_{Th}(z_j, \theta)) \]
\end{itemize}

We implemented these same $\chi^2$ expressions within our Markov Chain Monte Carlo (MCMC) analysis. For the BAO analysis, we use several observables. These include the rescaled angle-averaged distance $\mu = D_V(z)/(r_d \sqrt{z})$, the transverse comoving distance $z D_M(z)/(r_d \sqrt{z})$, and the Hubble horizon distance $D_H(z)/(r_d \sqrt{z})$. Here, $r_d$ represents the comoving sound horizon. For the Type Ia Supernovae (SNeIa) analysis, we use the distance modulus $\mu = m_b - M$ and the apparent magnitude $m_b$. The term $C^{-1}$ denotes the inverse of the covariance matrix for the respective dataset(BAO or SNeIa). Statistically, the $\chi^2$ function attains its minimum value at $\chi^2_{\text{min}} = \chi^2(\theta_{\text{best-fit}})$, where $\theta_{\text{best-fit}}$ represents the vector of parameters that best describe the observed data. 

For our proposed model, the parameter space is defined as $\theta_{\text{model}} = \{H_0, \Omega_0, \sigma\}$, while for the reference $\Lambda$CDM model, it is $\theta_{\Lambda\text{CDM}} = \{H_0, \Omega_m\}$. In this framework, if $\chi^2_{\text{min, model}} < \chi^2_{\text{min, } \Lambda\text{CDM}}$, the proposed model is statistically supported by the data. Conversely, if $\chi^2_{\text{min, model}} > \chi^2_{\text{min, } \Lambda\text{CDM}}$, the model is considered disfavored relative to the standard paradigm. The detailed results of the $\chi^2_{\text{min}}$ values for both models across all dataset combinations are presented in Table~ \ref{tab:full_cosmo_comparison}.

\begin{table*}[ht]
\centering
\small 
\begin{tabular}{|l|l|c|c|c|c|c|c|c|c|}
\hline
\textbf{Dataset Combination} & \textbf{Model} & $\chi^2_{OHD}$ & $\Delta\chi^2_{H}$ & $\chi^2_{BAO}$ & $\Delta\chi^2_{B}$ & $\chi^2_{SNeIa}$ & $\Delta\chi^2_{S}$ &  $\chi^2_{\text{total}}$ & $\Delta \chi^2$ \\ \hline

\multirow{2}{*}{OHD + Pantheon+} & $\Lambda$CDM & 70 & 0 & --- & --- & 1442 & 0 & 1512 & 0 \\ \cline{2-10} 
 & Model & 200 & 130 & --- & --- & 1414 & {\color{teal}-28} & 1613 & 101 \\ \hline

\multirow{2}{*}{OHD + DESI DR1 + Pantheon+} & $\Lambda$CDM & 69 & 0 & 44 & 0 & 1443 & 0 & 1556 & 0 \\ \cline{2-10} 
 & Model & 215  & 146 & 212 & 168 & 1423 & {\color{teal}-20} & 1850 & 294 \\ \hline

\multirow{2}{*}{OHD + DESI DR2 + Pantheon+} & $\Lambda$CDM & 67 & 0 & 102 & 0 & 1453 & 0 & 1622 & 0 \\ \cline{2-10} 
 & Model & 209 & 142 & 505 & 403 & 1471 & 18 & 2185 & 563 \\ \hline

\multirow{2}{*}{OHD + Union3} & $\Lambda$CDM & 45 & 0 & --- & --- & 34 & 0 & 79 & 0 \\ \cline{2-10} 
 & Model & 56 & 11 & --- & --- & 41 & 7 & 97 & 18 \\ \hline

\multirow{2}{*}{OHD + DESI DR1 + Union3} & $\Lambda$CDM & 48 & 0 & 14 & 0 & 31 & 0 & 93 & 0 \\ \cline{2-10} 
 & Model & 59 & 11 & 39 & 25 & 51 & 20 & 149 & 56 \\ \hline

\multirow{2}{*}{OHD + DESI DR2 + Union3} & $\Lambda$CDM & 50 & 0 & 11 & 0 & 31 & 0 & 92 & 0 \\ \cline{2-10} 
 & Model & 64 & 14 & 58 & 47 & 54 & 23 & 176 & 84 \\ \hline

\multirow{2}{*}{OHD + DESY5} & $\Lambda$CDM & 55 & 0 & --- & --- & 1662 & 0 & 1717 & 0 \\ \cline{2-10} 
 & Model & 114 & 59 & --- & --- & 1643 & {\color{teal}-19} & 1757 & 40 \\ \hline

\multirow{2}{*}{OHD + DESI DR1 + DESY5} & $\Lambda$CDM & 54 & 0 & 15 & 0 & 1663 & 0 & 1732 & 0 \\ \cline{2-10} 
 & Model & 117 & 63 & 88 & 73 & 1659 & {\color{teal}-4} & 1874 & 132 \\ \hline

\multirow{2}{*}{OHD + DESI DR2 + DESY5} & $\Lambda$CDM & 53 & 0 & 19 & 0 & 1665 & 0 & 1737 & 0 \\ \cline{2-10} 
 & Model & 115 & 62 & 173 & 154 & 1685 & 20 & 1973 & 236 \\ \hline
\end{tabular}

\vspace{0.3cm}
\caption{Comprehensive statistical breakdown of $\chi^2_{\text{min}}$ and sectoral $\Delta\chi^2_{\text{min}}$ values. We compare the performance of the $\Lambda$CDM model against the proposed model across nine distinct dataset combinations, including OHD, BAO, and SNeIa (DESY5). The relative statistical merit is quantified by $\Delta\chi^2 = \chi^2_{\text{Model,total}} - \chi^2_{\Lambda\text{CDM,total}}$, where $\Delta\chi^2_{H}$, $\Delta\chi^2_{B}$, and $\Delta\chi^2_{S}$ represent the individual contributions from the Hubble, BAO, and Supernovae sectors, respectively. }
\label{tab:full_cosmo_comparison}
\end{table*}

Upon inspecting Table~\ref{tab:full_cosmo_comparison}, it becomes evident that the proposed model is significantly disfavored and incurs a substantial statistical penalty when compared to the $\Lambda$CDM baseline. The model yields a considerably higher total $\chi^2_{\text{min}}$, indicating that $\Lambda$CDM maintains a dominant statistical preference in this analysis.

However, this pronounced discrepancy in $\chi^2_{\text{min}}$ values warrants a detailed explanation. While $\chi^2_{\text{min}}$ analysis primarily assesses the goodness-of-fit, the results are inherently influenced by the structural complexity of the model itself. To understand this, we must first examine the functional form of our model:
$$H^2(z) = H^2_{0}[\frac{\Omega_{0}}{1-\sigma}(1+z)^{(3+\sigma)} + (1- \frac{\Omega_{0}}{1-\sigma})(1+z)^{(2\sigma+2)}]$$
where $\Omega_0$ represents the background matter density parameter. In contrast, the standard $\Lambda$CDM model is defined as:
$$H^2(z) = H^2_0\left[ \Omega_{m}(1+z)^3 +(1 -\Omega_{m})\right]$$
The $\Lambda$CDM model possesses one fewer degree of freedom and exhibits greater structural symmetry. The fundamental differences in our model arise from the emergence of an effective matter density, $\Omega_{0,\text{eff}} = \Omega_0/(1-\sigma)$, and the modified evolution of the density components, which behave as $(1+z)^{(3 +\sigma)}$ and $(1+z)^{(2+2\sigma)}$. 

This increased complexity leads to a significant shift in the best-fit parameter space, forcing $\Omega_0$ toward values in the range of $\approx 0.6$ to $0.8$. Such values represent a drastic departure from the standard $\Lambda$CDM results. We have further explored this behavior by analyzing the variation of $\chi^2$ as a function of $\Omega_0$ and $\Omega_{m}$ to illustrate why the model struggles to achieve a competitive fit.

\begin{figure*}[t]
\centering

\begin{tabular}{cc}
    \includegraphics[width=0.48\textwidth]{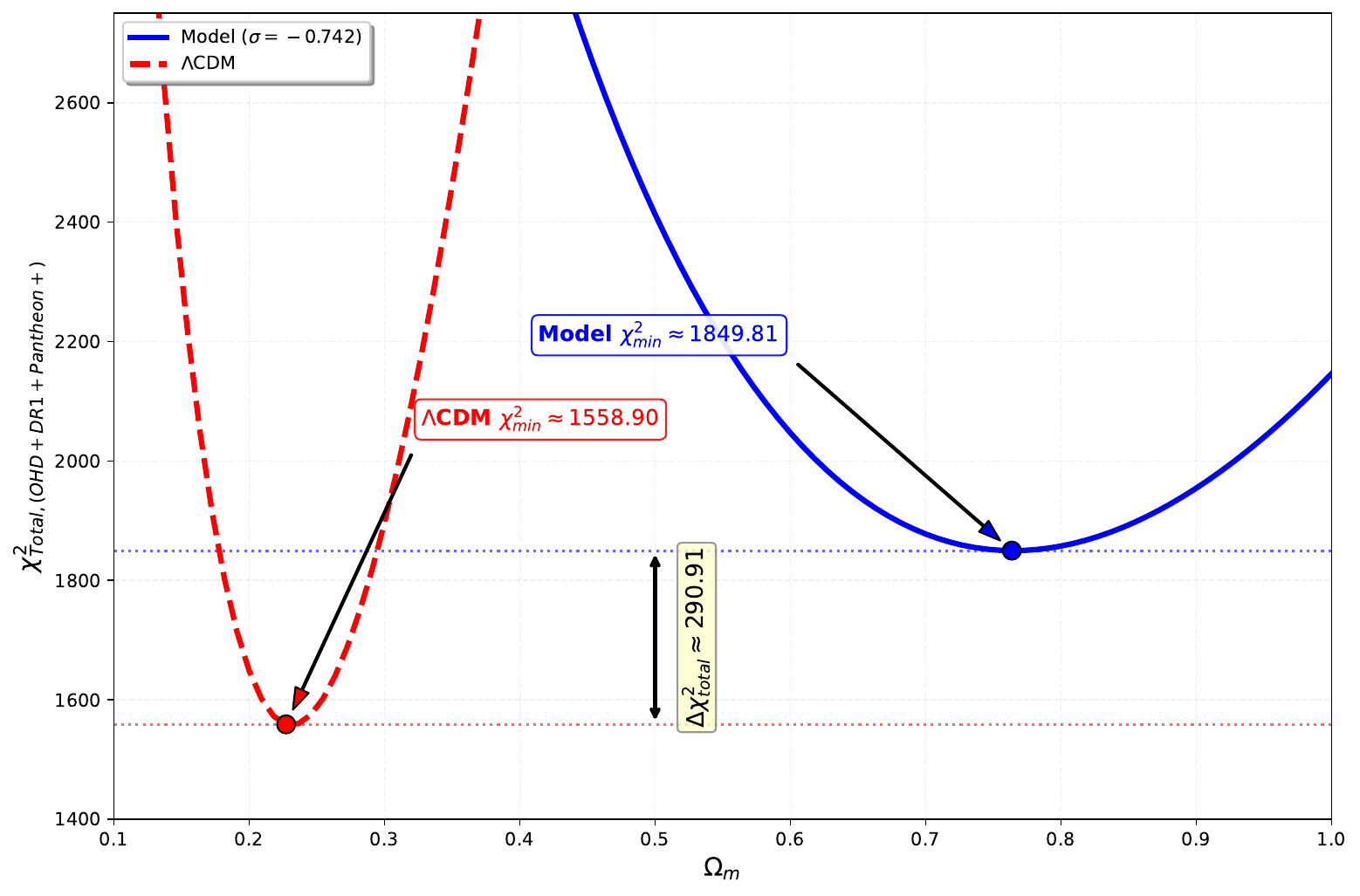} &
    \includegraphics[width=0.48\textwidth]{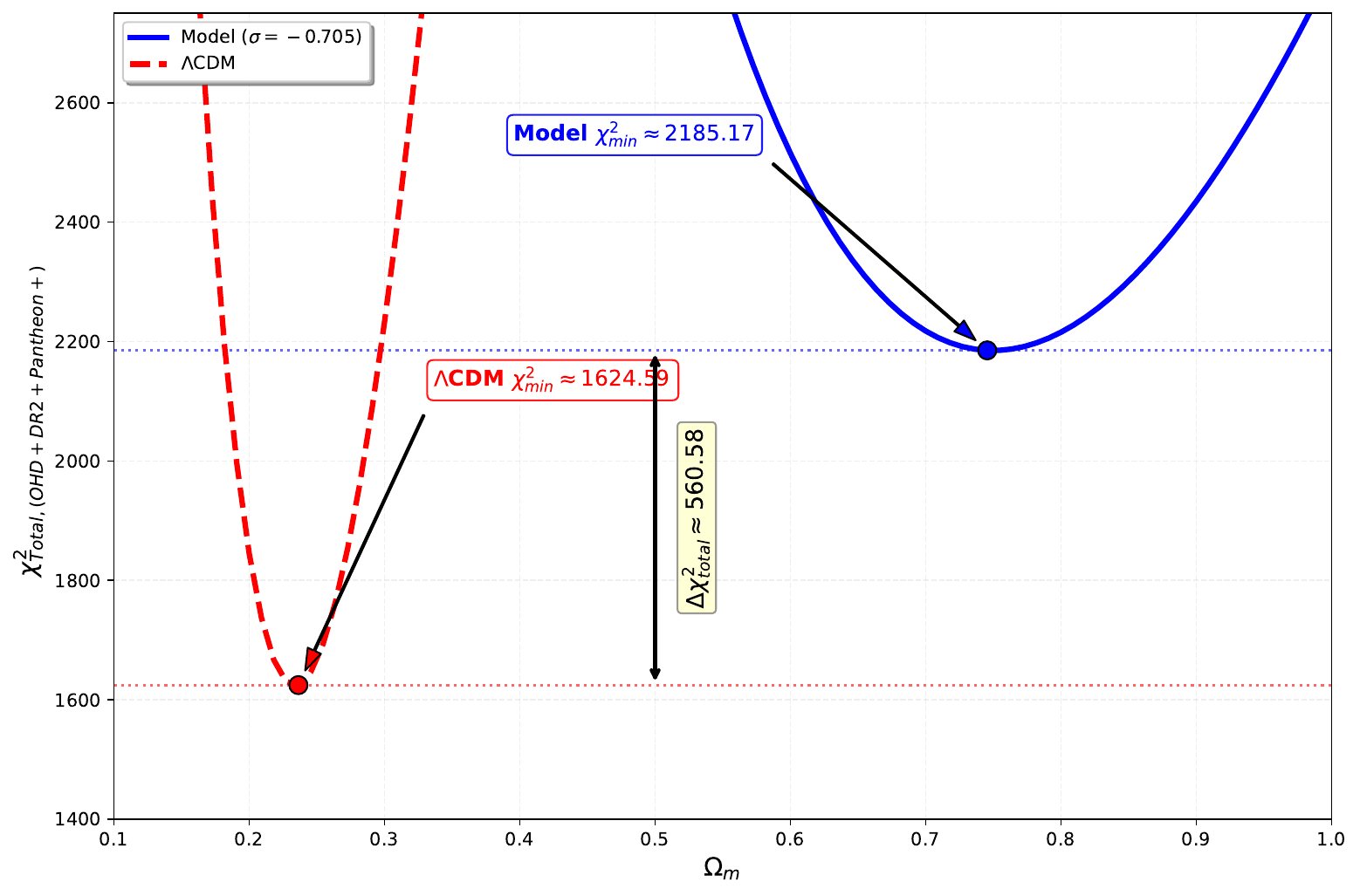} \\

    \small (a) OHD+DESI DR1+Pantheon+ &
    \small (b) OHD+DESI DR2+Pantheon+ \\[0.4cm]

    \includegraphics[width=0.48\textwidth]{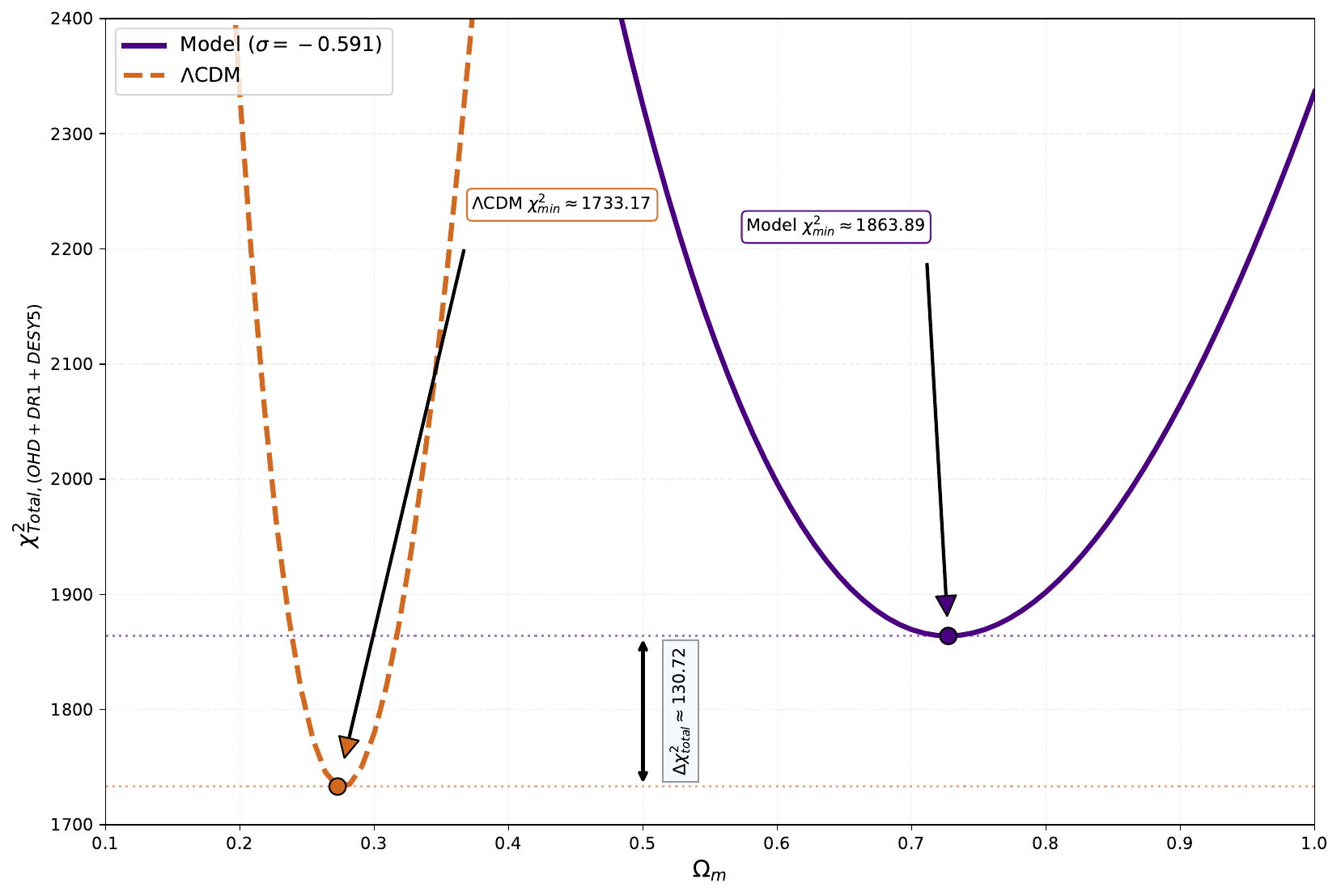} &
    \includegraphics[width=0.48\textwidth]{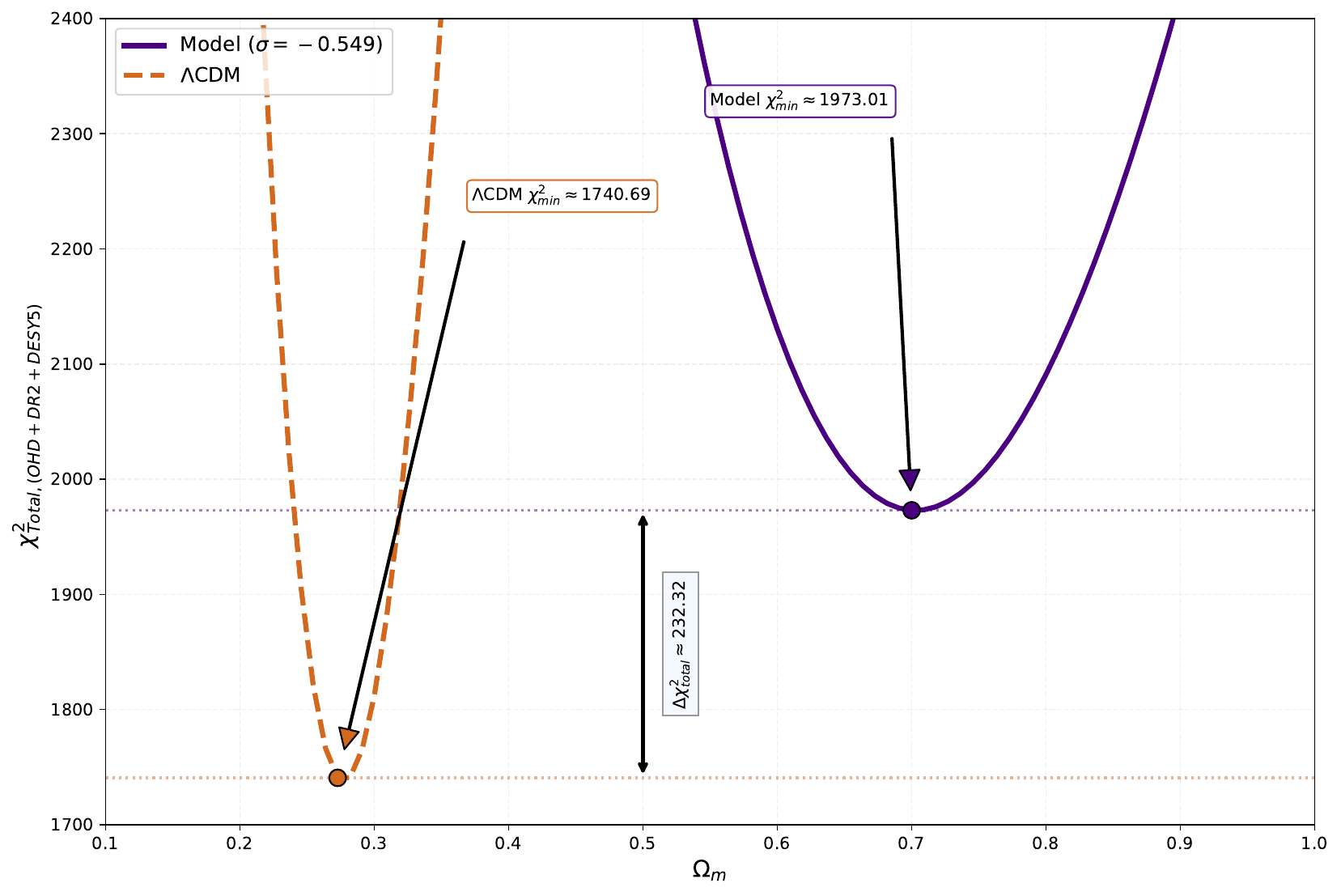} \\

    \small (c) OHD+DESI DR1+DESY5 &
    \small (d) OHD+DESI DR2+DESY%
\end{tabular}

\caption{Comparative analysis of $\chi^2_{\text{total}}$ as a function of $\Omega_m$ for $\Lambda$CDM and the proposed model. 
Top row uses DESY5 samples; bottom row uses Pantheon+ data.}
\label{fig:full_cosmo_grid}

\end{figure*}

In Fig. \ref{fig:full_cosmo_grid}, we illustrate the behavior of $\chi^2$ as a function of $\Omega_m$ for the joint analysis of BAO data with DESY5 and Pantheon+. 

By visualizing Fig. \ref{fig:full_cosmo_grid}, it becomes evident that our model yields a minimum $\chi^2$ at relatively large values of $\Omega_m$ (specifically the best-fit values in the range of $\approx 0.7$ to $0.8$). The distinction between our model and $\Lambda$CDM is clearly observable; therefore, we have presented the $\chi^2_{\text{min}}$ values for each data subsection. 

As shown in Table \ref{tab:full_cosmo_comparison}, the inclusion of DESI data results in a significant discrepancy in $\chi^2_{\text{min}}$ values, with DR2 providing a larger difference than DR1. Additionally, the OHD data provides a meaningful contribution to this analysis. However, when examining Pantheon+ and DESY5 individually, we observe that they signal lower $\chi^2_{\text{min}}$ values compared to $\Lambda$CDM.

\begin{figure*}[t]
\centering

\begin{tabular}{cc}
    \includegraphics[width=0.48\textwidth]{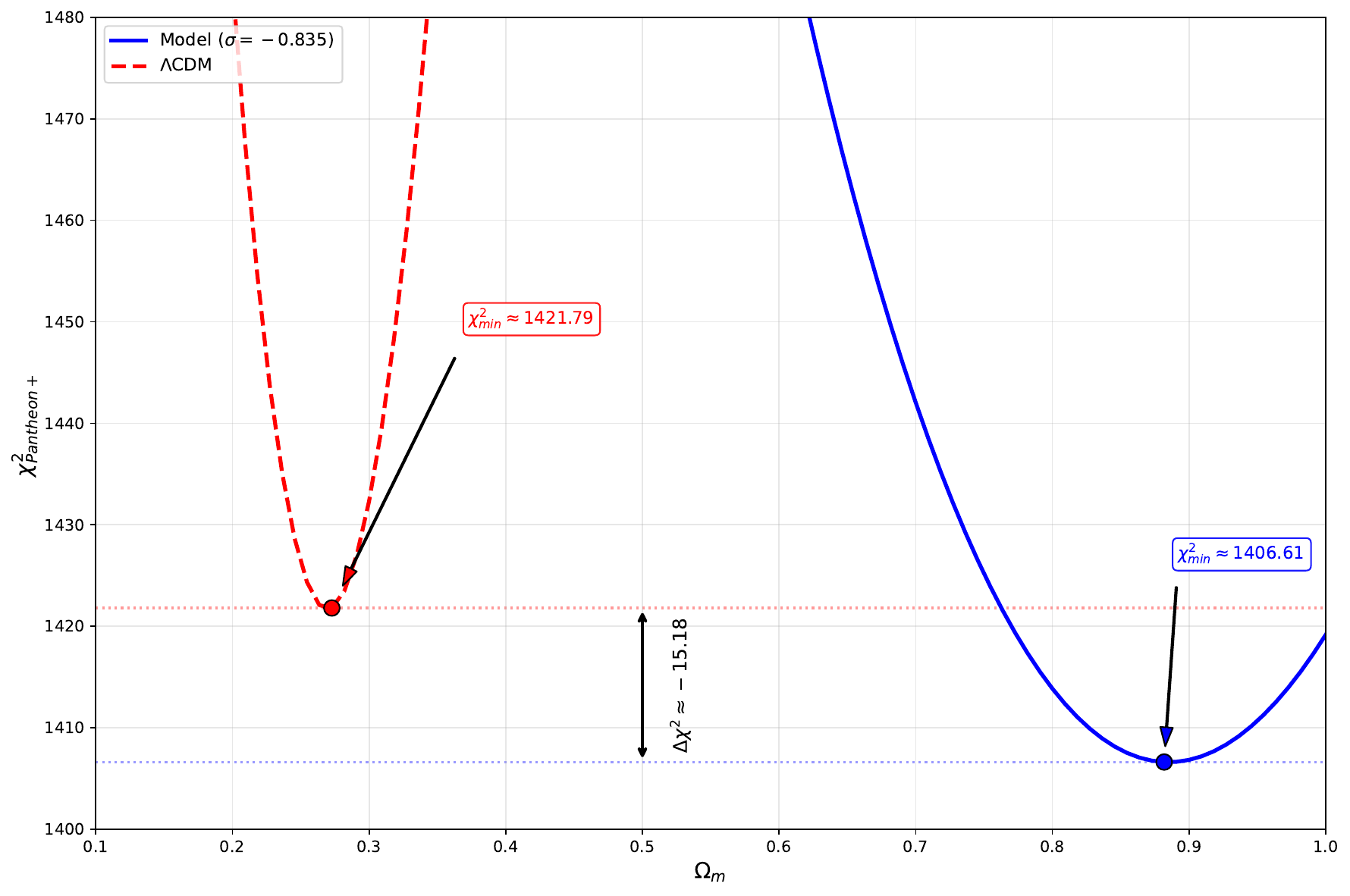} &
    \includegraphics[width=0.48\textwidth]{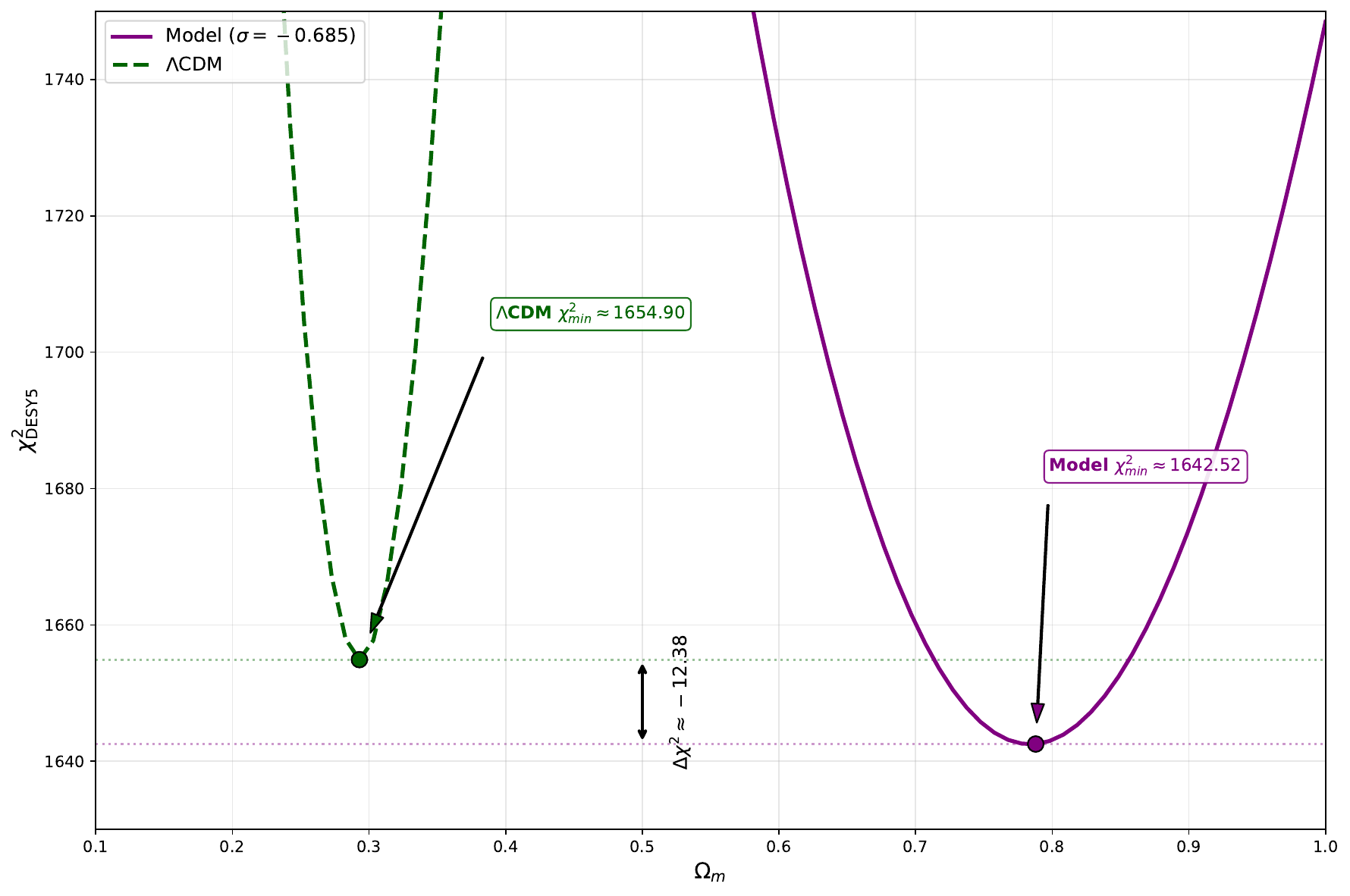} \\

    \small (a) OHD+Pantheon+ &
    \small (b) OHD+DESY5

\end{tabular}
\caption{Individual $\chi^2_{\text{SNeIa}}$ distributions versus $\Omega_{m}$ for the Pantheon+ and DESY5 supernova datasets. }
\label{chi-sn}
\end{figure*}

We have represented these findings graphically in Fig. \ref{chi-sn}, where a potentially lower $\chi^2_{\text{min}}$ value for our model is visible. Notably, this statistical advantage disappears upon the inclusion of DESI DR2. This suggests that a statistical gain for our model is only achievable when using Pantheon+ or DESY5 data in isolation. In these specific cases, the minimum $\chi^2$ tends toward $\Omega_m > 0.8$. 

According to this results, the majority of the cosmic density is composed of matter, while the remaining smaller portion is attributed to the modified curvature term, which scales as $(1+z)^{2+2\sigma}$. We have highlighted these results in Table \ref{tab:full_cosmo_comparison} in green. Consequently, in the case of joint analysis, $\Lambda$CDM remains statistically highly favored. In this context, our model is not currently competitive with $\Lambda$CDM; however, our objective is to demonstrate an alternative mechanism for producing cosmic acceleration without the need for any hypothetical fluids.

\section{Summary and Conclusions}
\label{sum}
The model presented in Equation~\eqref{Hubble} focuses on the special degrees of freedom characterized by the parameter $\sigma$. This parameter originates from the extended Non-Commutative (NC) fluid algebra and serves as the primary deformation variable in our framework.

The presence of $\sigma$ leads to modifications in both the effective equation of state and the density evolution. In the absence of this parameter ($\sigma=0$), the universe would consist solely of cold dark matter and curvature, which are insufficient to drive late-time acceleration. Therefore, the modification introduced by the NC term plays a crucial role in the background cosmology. Notably, it eliminates the requirement for a cosmological constant or an additional dark energy fluid to explain the current accelerated expansion. This is supported by both theoretical and observational analyses, where observational constraints are found to be consistent with the theoretical bounds on $\sigma$. Consequently, this model offers a compelling alternative explanation for the accelerating universe without invoking the concept of ``dark energy.''

Another significant feature of this model is the non-conservation of background matter, described by the continuity equation:
\begin{equation}
    \dot{\rho}_0 + 3H \rho_0 = \Gamma,
    \label{cc}
\end{equation}
where in our framework, $\Gamma = -\sigma H \rho_0$. Since $\Gamma \neq 0$, this term mimics a matter-creation scenario or an internal energy transfer within the system.

Theoretical considerations and observational constraints indicate that $\Gamma > 0$. While $H$ and $\rho_0$ are strictly positive, observational fits and theoretical modeling suggest that the parameter $\sigma$ is negative ($\sigma < 0$). Thus, the product $\Gamma = -\sigma H \rho_0$ yields a positive value. This highlights a matter-creation-like picture; however, as discussed previously, we refrain from claiming explicit particle production, interpreting it instead as a geometric effect. 
The key findings and future directions of this work are summarized as follows:

\begin{itemize}
    \item We have developed a cosmological model by implementing Non-Commutative (NC) geometry within a Newtonian fluid dynamics framework. A significant advantage of this model is its ability to exhibit cosmic acceleration without invoking hypothetical entities such as dark energy.
    
    \item The central feature of this model is the NC-originated parameter $\sigma$. Although derived from the underlying non-commutative theory, $\sigma$ serves as a measurable quantity that can be tested against current observational datasets.
    
    \begin{center}
    
    \end{center}

    \item Theoretical and observational analyses both support a viable range of $-1 < \sigma < 0$ to explain the observed late-time acceleration. Studies of energy conditions further validate this range as physically consistent.

    \item While Newtonian cosmology has inherent limitations—such as its inability to fully describe the radiation-dominated era, CMB physics, or super-horizon scales—it remains a robust and tractable framework for studying late-time cosmic evolution. Because of these limits, it is difficult to explain the inflationary era or early universe physics. Our model focuses only on late-time evolution and the accelerating universe.

    \item Although the model is statistically less favored than the standard $\Lambda$CDM model, it is noteworthy that recent SNeIa datasets, specifically Pantheon+ and DES-Y5, show a promising degree of favorability for this NC-based approach.

    \item The model leads to an apparent matter-creation scenario. However, in the absence of an explicit physical source, we interpret this phenomenon as a geometric effect arising from the deformed phase-space measure rather than physical particle production.

    \item While a full CMB-based analysis is currently outside the scope of this Newtonian framework, the model provides a foundation for performing growth-of-structure analyses using Newtonian perturbation theory.

    \item Furthermore, there is significant potential to explore the stability and evolution of this model using the tools of dynamical systems analysis.
    
     \item While the present study focuses on background dynamics, it is worth noting that in standard Newtonian perturbation theory, the 'Source' term(growth equation:$\ddot{\delta} + \text{(Friction)}\dot{\delta} + (\text{Source} ) \delta = 0$) in the growth equation is purely gravitational. In our framework, the non-commutative parameter $\sigma$ contributes an additional effective term to this dynamics. Investigating how this modifies the growth of structures would impose further constraints on the model and represents a compelling direction for future research.
\end{itemize}

\section{Acknowledgmebt}
We are thankful to Prof. Subir Ghosh from ISI, Kolkata,  Dr. Supriya Pan and Sivasish Paul from Presidency University, Kolkata and Dr Souvik Ghose from HRI, Allahabad, for their helpful discussions. 
Raj Kumar Das is also thankful to Rahul Shah from ISI, Kolkata, for some helpful discussions.
We thank ARIES and Presidency University for providing the necessary computational facility. { We are also sincerely grateful to the editorial team and the anonymous reviewers for their valuable suggestions and for granting an extension of the revision deadline.}
\newpage

\clearpage

\bibliography{biblio}

\end{document}